\documentclass[twocolumn,floats,floatfix,amssymb,nofootinbib,prd,superscriptaddress]{revtex4-2}
\usepackage{graphicx}
\usepackage{amsmath,amssymb}
\usepackage{amsfonts}

\usepackage{xspace} 
\usepackage[usenames]{color}
\usepackage{dcolumn}
\usepackage{bm}
\usepackage{mathrsfs}
\usepackage[colorlinks=true]{hyperref}
\usepackage[all]{hypcap} 
\usepackage[utf8]{inputenc} 
\usepackage{multirow}
\usepackage[normalem]{ulem}

\usepackage{etoolbox}
\usepackage{tikz}

\newcommand{\ben}{\begin{enumerate}}
\newcommand{\een}{\end{enumerate}}

\def\be{\begin{equation}}
\def\ee{\end{equation}}
\newcommand{\beq}{\begin{eqnarray}}
\newcommand{\eeq}{\end{eqnarray}} 
\newcommand{\ba}{\begin{align}}
\newcommand{\ea}{\end{align}}

\def\be{\begin{equation}}
\def\ee{\end{equation}}
	
\newcommand{\bea}{\begin{eqnarray}}
\newcommand{\eea}{\end{eqnarray}}

\usepackage{physics}

\begin{document}

\title{On the instability of ultracompact horizonless spacetimes}

\author{Zhen Zhong}
\affiliation{CENTRA, Departamento de F\'{\i}sica, Instituto Superior T\'ecnico -- IST, Universidade de Lisboa -- UL,
Avenida Rovisco Pais 1, 1049-001 Lisboa, Portugal}
\author{Vitor Cardoso}
\affiliation{Niels Bohr International Academy, Niels Bohr Institute, Blegdamsvej 17, 2100 Copenhagen, Denmark}
\affiliation{CENTRA, Departamento de F\'{\i}sica, Instituto Superior T\'ecnico -- IST, Universidade de Lisboa -- UL,
Avenida Rovisco Pais 1, 1049-001 Lisboa, Portugal}
\author{Elisa Maggio}
\affiliation{Max Planck Institute for Gravitational Physics (Albert Einstein Institute)
Am Muhlenberg 1, 14476 Potsdam, Germany}
\begin{abstract}
Motivated by recent results reporting the instability of horizonless objects with stable light rings, we revisit the linearized stability of such structures. 
In particular, we consider an exterior Kerr spacetime truncated at a surface where Dirichlet conditions on a massless scalar are imposed. 
This spacetime has ergoregions and light rings when the surface is placed sufficiently deep in the gravitational potential. We establish that the spacetime is linearly, mode-unstable when it is sufficiently compact, in a mechanism associated with the ergoregion. In particular, such instability has associated zero-modes. At large multipole number the critical surface location for zero modes to exist is {\it precisely} the location of the ergosurface along the equator. We show that such modes don't exist when the surface is outside the ergoregion, and that any putative linear instability mechanism acts on timescales $\tau \gtrsim 10^5 M$, where $M$ is the black hole mass. Our results indicate therefore that at least certain classes of objects are linearly stable in the absence of ergoregions, even if rotation and light rings are present. 
\end{abstract}

\maketitle
\section{Introduction}
The special properties of black hole (BH) horizons, and the failure of classical General Relativity in their interior calls for outstanding observational evidence for BHs~\cite{Abramowicz:2002vt,Cardoso:2019rvt}. In parallel, theoretical arguments constraining the universe of alternatives are welcome. Stability arguments are a robust indicator for the feasibility of the equilibrium solutions of a given theory. In fact, the very existence of structure -- galaxies, planets, stars -- is due to a wide array of instability mechanisms, such as Jeans'~\cite{Jeans:1902fpv,Chandra:1998nia}. In the context of the gravitational physics of very compact objects, two mechanisms can play a role, and they are tied to the distinctive features
of horizons, or the absence thereof. These mechanisms hinge on fundamental aspects of General Relativity, specifically the existence of ergoregions and regions of the spacetime where lensing is so strong that photon orbits can ``close,'' and which therefore work as trapping regions.

The vacuum Kerr spacetime possesses an {\it ergoregion}, a region within which static timelike observers don't exist and ``negative energy'' states are allowed. The existence of the ergoregion allows for efficient extraction of energy from spinning BHs~\cite{Penrose:1971uk,zeldovich1,zeldovich2,Blandford:1977ds,Brito:2015oca}. In the absence of horizons, ergoregions give rise to a linear instability: any small negative-energy fluctuation within the ergoregion must trigger a positive-energy state upon traveling to the exterior of the ergoregion (where only positive-energy states are allowed). Energy conservation then implies that the negative energy states inside must grow in amplitude, triggering an exponentially growing cascade~\cite{Friedman:1978wla,1978CMaPh..63..243F,CominsSchutz,Moschidis:2016zjy,Vicente:2018mxl,Brito:2015oca}. This mechanism was shown to be effective for spinning compact objects, with timescales which are astrophysically relevant~\cite{1996MNRAS.282..580Y,Kokkotas:2002sf,Cardoso:2005gj,Cardoso:2007az,Cardoso:2008kj,Chirenti:2008pf,Pani:2010jz,Hod:2017wks,Maggio:2017ivp,Maggio:2018ivz}. Thus, spinning objects whose exterior is close to Kerr, but which do not have horizons should be spinning down and emitting copious amounts of gravitational waves. A stochastic gravitational-wave background from spin loss has not been detected yet, thus excluding classes of horizonless compact objects via observations~\cite{Barausse:2018vdb}.

In addition to the ergoregion instability, it was argued that even non-spinning objects should be unstable, if compact enough to develop light rings, against a nonlinear mechanism. Schwarzschild BHs have a single, unstable photon surface. However, in the absence of horizons, stable photon surfaces necessarily appear~\cite{Keir:2014oka,Cardoso:2014sna,Cunha:2017qtt}. In these spacetimes, linearized fluctuations decay extremely slowly, leading to the conjecture that nonlinear effects might cause either a collapse to a BH or dispersion of star material~\cite{Keir:2014oka,Cardoso:2014sna}. Unlike the ergoregion instability, this ``trapping instability'' (TI) is nonlinear in nature. Therefore, estimation of timescales or even the verification that the instability is present is a formidable problem.

Nevertheless, it was recently reported that the TI was observed in two different classes of objects made of fundamental fields, i.e. boson and Proca stars~\cite{Cunha:2022gde}. The objects all
were spinning stars and the instability timescale was always relatively short, raising the possibility that the mechanism observed is not a TI in nature, and possibly not even nonlinear, but rather something else (for example, as we discuss in the main text, the stiffness of the system under study could introduce artificial effects). The work reported in Ref.~\cite{Cunha:2022gde} motivated us to understand in finer detail the ergoregion instability of Kerr-like objects, studied in the literature but not exhaustively
~\cite{Cardoso:2008kj,Maggio:2017ivp,Maggio:2018ivz,Cardoso:2019rvt}. In particular, Refs.~\cite{Maggio:2017ivp,Maggio:2018ivz} studied  the instability when the surface sits deep in the gravitational well (in fact, close to the horizon). A study on the ergoregion instability threshold was done in a fluid setup~\cite{Oliveira:2014oja}, where strong evidence was found that the critical surface is indeed the ergosurface (see also Ref.~\cite{Hod:2017eld} where zero modes -- a property of special interest in Kerr-like geometries, as we will show -- were investigated).

Here, we aim to explore further the ergoregion instability in the exterior Kerr spacetime (truncated at a finite radius outside the horizon), and to understand at which surface the instability is quenched (do we find numerical evidence that it coincides with the ergosurface?) and whether new (linear) instabilities -- related to the presence of light rings -- set in even when the surface sits outside the ergosurface. We note that the interplay between ergoregions and light rings is made all the more interesting since stationary, axisymmetric, and asymptotically flat spacetime in $1 + 3$ dimensions with an ergoregion must have at least one light ring on its exterior~\cite{Ghosh:2021txu}.

\section{Setup}
\subsection{The spacetime, coordinates and dynamical equations}
In Boyer-Lindquist coordinates, the metric of Kerr spacetime can be written as
\beq
d s^{2}&=&\left(1-\frac{2 M r}{\Upsilon^{2}}\right) d t^{2}+\frac{4 a M r \sin ^{2} \theta}{\Upsilon^{2}} d t d \varphi\nonumber\\
&-&\left[\left(r^{2}+a^{2}\right) \sin^{2}\theta+\frac{2 M r}{\Upsilon^{2}} a^{2} \sin^{4} \theta\right] d \varphi^{2}\nonumber\\
&-&\frac{\Upsilon^{2}}{\Delta} d r^{2}-\Upsilon^{2} d \theta^{2} \,,\label{kerr:geometry}
\eeq
where
\begin{equation}
  \Delta=r^{2}-2 M r+a^{2}\,, \qquad \Upsilon^{2}=r^{2}+a^{2} \cos^{2} \theta\,.
\end{equation}
The horizons of this geometry are located at $r_{\pm} = M \pm \sqrt{M^2 - a^2}$, i.e. the outer event and Cauchy horizons, respectively. These will be absent in our construction.
The ergosurface is defined by the zeros of $g_{tt}$, which is the torus $r_{\rm ergo}=M+\sqrt{M^2-a^2\cos^2\theta}$. On the equator and the poles, $r_{\rm ergo}=2M,\,M+\sqrt{M^2-a^2}$, respectively. The ergoregion is the chief responsible for a linear instability, which is governed by the angular velocity
\be
\Omega=\frac{a}{2Mr_+}\,.
\ee
The spacetime also has unstable light rings at $r_{\rm LR}=2M\left(1+\cos\left(\frac{2}{3}\arccos(\mp a/M)\right)\right)$. When $a=\sqrt{2}M/2\sim 0.707$, the co-rotating light ring sits at the same radius than the ergoregion, in the equatorial plane ($r=2M$).

On the Kerr background, a single master equation governs perturbations of massless fields $\Psi$ \cite{Teukolsky:1973ha}
\begin{equation}
\begin{gathered}
\left[\frac{\left(r^2+a^2\right)^2}{\Delta}-a^2 \sin ^2 \theta\right] \partial_t^2 \Psi-\Delta^{-s} \partial_r\left(\Delta^{s+1} \partial_r \Psi\right) \\
+\frac{4 M a r}{\Delta} \partial_t \partial_{\varphi} \Psi+\frac{a^2}{\Delta} \partial_{\varphi}^2 \Psi- \mathcal{D}_s \Psi -2 s \frac{a(r-M)}{\Delta} \partial_{\varphi} \Psi\\
-2 s\left[\frac{M\left(r^2-a^2\right)}{\Delta}-r-i a \cos \theta\right] \partial_t \Psi=0\label{eq:Teukolsky}\,,
\end{gathered}
\end{equation}
where $\Psi$ is a field with spin $s$, $\mathcal{D}_s$ is the spin-weighted spherical Laplacian given by
\begin{equation}
 \mathcal{D}_s \equiv \frac{1}{\sin \theta} \pdv{\theta}\left(\sin \theta \pdv{\theta}\right)+\left(s-\frac{\left(-i\partial_{\varphi}+s \cos \theta\right)^2}{\sin ^2 \theta}\right)\,.\nonumber
\end{equation}

We will focus for simplicity on scalar fields, $s=0$. We find no reason why scalars should have special properties in this context, so we expect similar results for other massless fields.
Following Ref.~\cite{Ripley:2022ypi}, we briefly review and introduce the horizon penetrating, hyperboloidally compactified coordinates $\{\tau,\rho,\theta,\phi\}$, which is a natural choice for studying BH perturbations~\cite{Zenginoglu:2011jz}. First of all, the ingoing coordinates $\{v, r, \theta, \phi\}$ are given by
\begin{equation}
  dv=dt + \frac{2 M r}{\Delta}dr, \quad d\phi=d\varphi + \frac{a}{\Delta}dr\,.
\end{equation}
The hyperboloidal time variable $\tau$ is defined as
%
\be
\dd{\tau} \equiv \dd{v} -\left(1+\frac{4M}{r}\right)\dd{r}\,,
\ee
and the compactified radial coordinate $\rho$ as 
\be
\rho \equiv \frac{1}{r}\,.
\ee
By applying the separation of variables, setting
\begin{equation}
\Psi(\tau, \rho, \theta, \phi)=e^{-i \omega \tau} e^{i m \phi} S(\theta) R(\rho)\,,
\end{equation}
Eq.~\eqref{eq:Teukolsky} is separated into two equations
\beq
\nonumber &&\mathcal{C}_2 R''(\rho)+\mathcal{C}_1 R'(\rho)+\left(\mathcal{C}_0-{}_0\mathcal{A}_{\ell m}(c)\right) R=0\,, \\ \\
&&\frac{1}{\sin \theta} \frac{d}{d \theta}\left(\sin \theta \frac{d S}{d \theta}\right)\nonumber\\
&+&\left(a^2 \omega^2 \cos ^2 \theta-\frac{m^2}{\sin ^2 \theta}+{}_0\mathcal{A}_{\ell m}(c)\right) S=0\,,
\eeq
where $c = a \omega$ is the oblateness parameter, ${}_0\mathcal{A}_{\ell m}(c)$ is the angular separation constant and
\begin{align}
&\mathcal{C}_2 = -\rho^2 (1-2 M \rho+a^2 \rho^2)\,, \\
&\mathcal{C}_1 = 2 i \omega-2\rho +2 i\left[am + a^2\omega - M (3i +8M\omega)\right] \rho^2 \nonumber\\
&\qquad +4 a^2 (2 i \omega M-1) \rho^3\,, \\
&\mathcal{C}_0 = \omega (2am + a^2\omega - 16M^2\omega) \nonumber\\
&\qquad + 2 (i +4\omega M)(am +a^2\omega - M (i +4M\omega)) \rho  \nonumber\\
&\qquad + 2a^2 (i+2\omega M)(i+4\omega M) \rho^2\,.
\end{align}
Notice that in the zero-rotation limit, ${}_0\mathcal{A}_{\ell m}(c)=\ell(\ell+1)$ where $\ell$ is the angular integer number
used to label the harmonics.
In this limit, scalar spheroidal harmonics are simply the standard spherical harmonics~\cite{Berti:2005gp}.
\subsection{Boundary conditions}
We will deal only with the exterior Kerr spacetime, by imposing boundary conditions at the surface of the ultracompact object, which we parametrize as
\be
r_0=r_+(1+\epsilon)\,.\label{eq:surface}
\ee
In particular, we enforce Dirichlet boundary conditions on the radial function, 
\be
R(1/r_0)=0\,,\label{eq_bc_definition}
\ee
which means that $\Psi$ also vanishes at the surface.

We will not deal with the interior region, and instead assume that it is composed of a material where condition \eqref{eq_bc_definition} holds.
Our main goal is to understand the ergoregion instability and possible linear instability mechanisms associated with the existence of light rings. 
Therefore, we assume that the matter content of the object is such that scalar waves are totally reflected~\footnote{This procedure is akin to studying reflection of electromagnetic waves off a perfect conductor~\cite{Jackson:1998nia}. In reality, a formal analysis would require one to provide the conductivity and permeability of the material, but the correct calculation in the perfect conductor limit allows one to simply impose boundary conditions at the surface of the conductor and to forget about the conductor itself.}. 
Dirichlet boundary conditions are also intended to mimic the regularity conditions in the interior of the object. 
We do not expect any new qualitative feature to arise from the introduction of the interior itself, 
but parameters describing the interior (e.g., the equation of state of matter, etc.) could possibly mask the physics we want to explore and bring in new unwanted complications. 
The physics we want to understand is related to features that are found in the exterior vacuum spacetime already.

The above setup has two necessary ingredients that we require: an ergoregion and a trapping region. There is no stable light ring, instead the trapping is caused by the Dirichlet conditions at the boundary, which confine perturbations in the region between the surface and the unstable light ring. This feature can be more easily seen in non-spinning geometries, which are governed by a wavelike equation with a potential peaked at close to the $r=3M$ surface~\cite{Berti:2009kk}. We have confirmed numerically that large $\ell$ modes are extremely long-lived when $a=0$, as in Ref.~\cite{Cardoso:2014sna}. 
Note also that the non-spinning geometry of compact stars is stable, hence putative new features -- if there are any at the linear order -- should be associated with {\it rotation and trapping}, both present in our setup.

However, it is still an interesting issue to understand at least the effects of allowing the scalar wave to probe the interior. With this in mind, we discuss a simple two-dimensional toy model in Appendix~\ref{appendix:interior}, where we implement superradiance via a Lorentz-violating term in the Klein-Gordon equation (following the original work by Zel'dovich~\cite{zeldovich1,zeldovich2}), and we mimic the star interior assigning an absorption parameter and a sound speed different from unity in its interior. Our main result from Appendix~\ref{appendix:interior} is Fig.~\ref{fig:2d_S1}, which shows that imposing Dirichlet boundary conditions at the surface is a limiting case of dealing with the interior of an absorbing star. Thus, the toy model further supports our use of a simple Kerr exterior where the boundary conditions are imposed at the cutoff radius.

\section{Numerical Approach}
\subsection{Frequency-domain calculations of the spectra}
A robust method to deal with the eigenfrequencies of Kerr BHs -- stable, well-tested and widely used in the computation of Kerr quasi-normal modes (QNMs) -- consists on a continued-fraction representation of the problem, also known as Leaver method~\cite{Leaver:1985ax, leaver1986solutions}. Unfortunately, the method is well suited for BH spacetimes, but not for the problem at hand, where we need to enforce boundary condition \eqref{eq_bc_definition} at a finite radius $r_0$.
We use instead the approach by Ripley~\cite{Ripley:2022ypi}, which discretizes the radial equation using a Chebyshev pseudospectral method, and use the Cook-Zalutskiy spectral approach~\cite{Cook:2014cta} to solve the angular sector.
The spin-weighted spheroidal harmonics are related to the angular spheroidal function of the first kind when $s = 0$ and $\varphi = 0$, see Refs.~\cite{flammer, Berti:2005gp,https://doi.org/10.1029/RS005i008p01207,https://doi.org/10.1111/j.0022-2526.2004.01526.x} for an extensive discussion.
For $c = 0$, ${}_0\mathcal{A}_{\ell m}(0)$ increases monotonically with $\ell$, but for $c \neq 0$, the $\ell$-th eigenvalue is not defined uniquely. In Ref.~\cite{Cook:2014cta}, the $\ell$-th eigenvalue is identified via continuity along some sequence of solutions connected to the well-defined value of ${}_0\mathcal{A}_{\ell m}(0)$. In our case, since we do not study extreme spin parameters, we find that the $\ell$-th smallest eigenvalue is same as previous one. For a more detailed explanation, please refer to Ref.~\cite{Cook:2014cta}.

Furthermore, due to the rapid change of the eigenfunction as the boundary approaches the horizon, we made the following improvements to Ripley's method. First of all, the matrix given by Chebyshev pseudospectral method with Dirichlet boundary condition \eqref{eq_bc_definition} is very ill-conditioned, so we replace the Chebyshev pseudospectral method with ultraspherical or Olver-Townsend spectral method~\cite{olver2013fast}, which gives sparser and better conditioned matrices and was incorporated into \texttt{chebfun} \cite{Driscoll2014} and \texttt{ApproxFun.jl} \cite{ApproxFun.jl-2014} package. Especially, in one dimensional cases, the condition number of the matrix is bounded by a constant \cite{olver2013fast}.
%
%
%

To further confirm the correctness of our results, we also use a direct integration method to check our results.
%
For given parameters, we need to search the dominant mode and overtones. To avoid missing the modes we are interested in, we use the global complex root and pole finding (GRPF) algorithm \cite{10.1145/2699457, 8457320, Gasdia19} to search multiple modes. In addition, we also perform an adaptive sampling in the complex plane region using \texttt{adaptive} package \cite{Nijholt2019} to be further sure we are not missing some related modes.

\subsection{Time-domain analysis}
To further validate our results in the frequency domain, we also perform a time domain analysis following using the $1+1$ approach of Ref.~\cite{Dolan:2012yt}. Our procedure is almost identical, except that we use Dirichlet boundary condition \eqref{eq_bc_definition} at the left boundary and an outgoing boundary condition at the right boundary (typical $r \sim 1000M$)~\cite{Dima:2020rzg}. 
The decomposition of the scalar field can be written as
\begin{equation}
\Psi=\sum_{j=|m|}^{\infty} \psi_j(r) Y_{j m}(\theta) \,,
\end{equation}
where $Y_{j m}(\theta)$ is the spherical harmonic of degree $j$ and order $m$. To specify initial data, we first define a tortoise coordinate as
\begin{equation}
\dv{r_*}{r}=\frac{r^2+a^2}{\Delta} \,,
\end{equation}
or, after fixing the constant of integration,
\begin{equation}
  r_*=r+\frac{2 M}{r_{+}-r_{-}}\left(r_{+} \ln \left|\frac{r-r_{+}}{2 M}\right|-r_{-} \ln \left|\frac{r-r_{-}}{2 M}\right|\right)\,.
\end{equation}
Our initial condition is a time-symmetric Gaussian,
\begin{equation}
\psi_{l=m}=\exp \left(-\frac{r_*-r_c}{2 \sigma^2}\right), \quad \psi_{l>m}=0=\partial_t\psi_l,
\end{equation}
with $r_c = 10 M$ and $\sigma = 2M$.
We extract the dominant modes from the time series data using the Prony method \cite{Berti:2007dg} and compare them with the results in the frequency domain. 
%

\paragraph*{Stiffness and numerical instability} 
For some surface locations, the time-domain evolutions can become very challenging: the modes are extremely long-lived and the system becomes stiff,
leading to the appearance of numerical instabilities. Unlike physical instabilities, which we find and discuss below when ergoregions are present, numerical instabilities
are not robust against grid settings and in particular when the resolution increases. Nevertheless, they are important to identify as they set a limit on the region of the parameter space we are able to probe. We can estimate the stiffness of our system since the signal (as we will see below in more detail) has the late-time form
\be
\psi_{\ell = m} (t, r) = \sum_{j = 0}^n e^{-i \omega_j t} c_j(r) \,,
\ee
where $\omega_0$ is the dominant mode and $\omega_j (j > 0)$ are the $j$-th overtones. Normally there are infinite terms, but in the following we ignore the terms with $c_j \ll 1$. Then the stiffness ratio at some fixed radius is given by~\cite{lambert1991numerical}
\begin{equation}
\mathcal{S} = \frac{\abs{\mathrm{Im}(\omega_n)}}{\abs{\mathrm{Im}(\omega_0)}} \,.
\end{equation}
The ratio above is a measure of stiffness of the system; differential equation systems with larger stiffness ratio can be considered more stiff. From the frequency domain data in Table~\ref{table:modes_2M}, we have that $\abs{\mathrm{Im}(\omega_0)} \ll 1$ and decreases as $\ell = m$ increases, thus stiffness becomes more and more important. Accordingly, evolving the system for large timescales is challenging. Thus, in this work we limit ourselves to excluding possible instabilities with timescales $\tau\lesssim 10^5 M$ only.
%
\begin{table}[th]
  \caption{QNMs for $a = 0.5M$, $\ell = m$ and a surface at $r_0 = 2M$. Notice that modes with very small imaginary parts may have considerable relative error.
  \label{table:modes_2M}}
  \begin{ruledtabular}
  \begin{tabular}{ccccc}
  $\ell = m$ & $M\omega$  & ${}_0\mathcal{A}_{\ell m}(c)$ \\
  \hline
  $1$  & $-0.338-7.47\times 10^{-2}i$ & $1.98 - 1.02 \times 10^{-2} i$ \\
  $2$  & $-0.444-2.24\times 10^{-2}i$ & $5.97 - 2.86 \times 10^{-3} i$ \\
  $3$  & $-0.532-3.44\times 10^{-3}i$ & $12.0 - 4.09 \times 10^{-4} i$ \\
  $4$  & $-0.600-1.79\times 10^{-4}i$ & $20.0 - 1.96 \times 10^{-5} i$ \\
  $5$  & $-0.652-4.06\times 10^{-6}i$ & $30.0 - 4.09 \times 10^{-7} i$ \\
  $6$  & $-0.696-5.62\times 10^{-8}i$ & $42.0 - 5.24 \times 10^{-9} i$ \\
  $7$  & $-0.734-4.95\times 10^{-10}i$ & $56.0 - 4.29 \times 10^{-11} i$ \\
 \end{tabular}
 \end{ruledtabular}
\end{table}
\begin{table}[th]
 \caption{Dominant unstable QNMs for $a = 0.99M$ and a surface at $\epsilon = 10^{-3}$. 
 \label{table:modes}}
 \begin{ruledtabular}
 \begin{tabular}{ccccc}
 $\ell = m$ & $M\omega$  & ${}_0\mathcal{A}_{\ell m}(c)$ \\
 \hline
 $1$  & $0.339+2.29\times 10^{-5}i$ & $1.64 - 2.61 \times 10^{-6}i$ \\
 $2$  & $0.757+1.16\times 10^{-5}i$ & $5.91 - 2.55 \times 10^{-6} i$ \\
 $3$  & $1.18+4.82\times 10^{-6}i$ & $11.8 - 1.29 \times 10^{-6} i$ \\
 $4$  & $1.60+1.84\times 10^{-6}i$ & $19.8 - 5.52 \times 10^{-7} i$ \\
 $5$  & $2.02+6.69\times 10^{-7}i$ & $29.7 - 2.17 \times 10^{-7} i$ \\
\end{tabular}
\end{ruledtabular}
\end{table}
%

\section{Results}
%
\begin{figure}[th]
\includegraphics[width=0.5\textwidth]{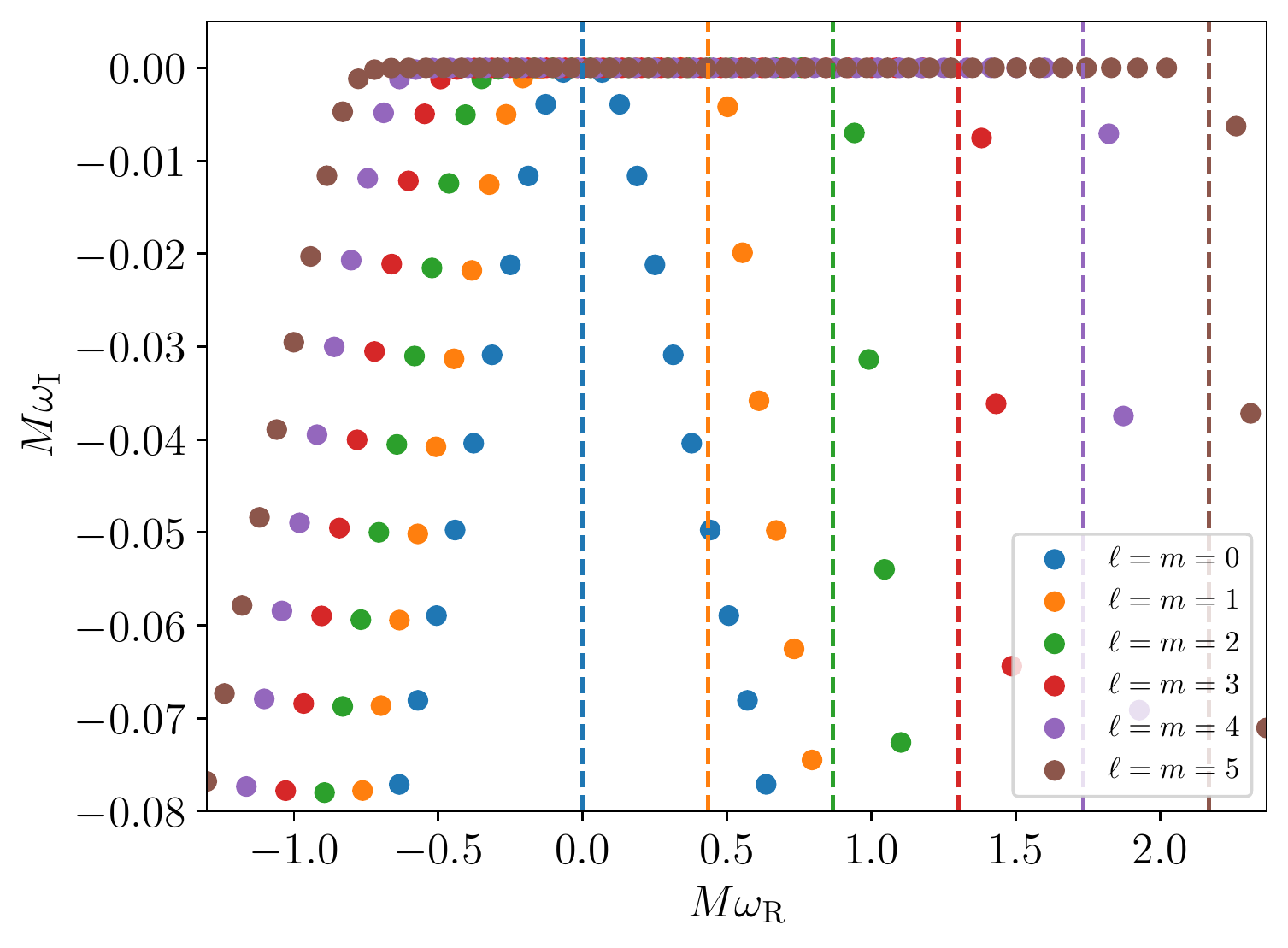}\\
\includegraphics[width=0.5\textwidth]{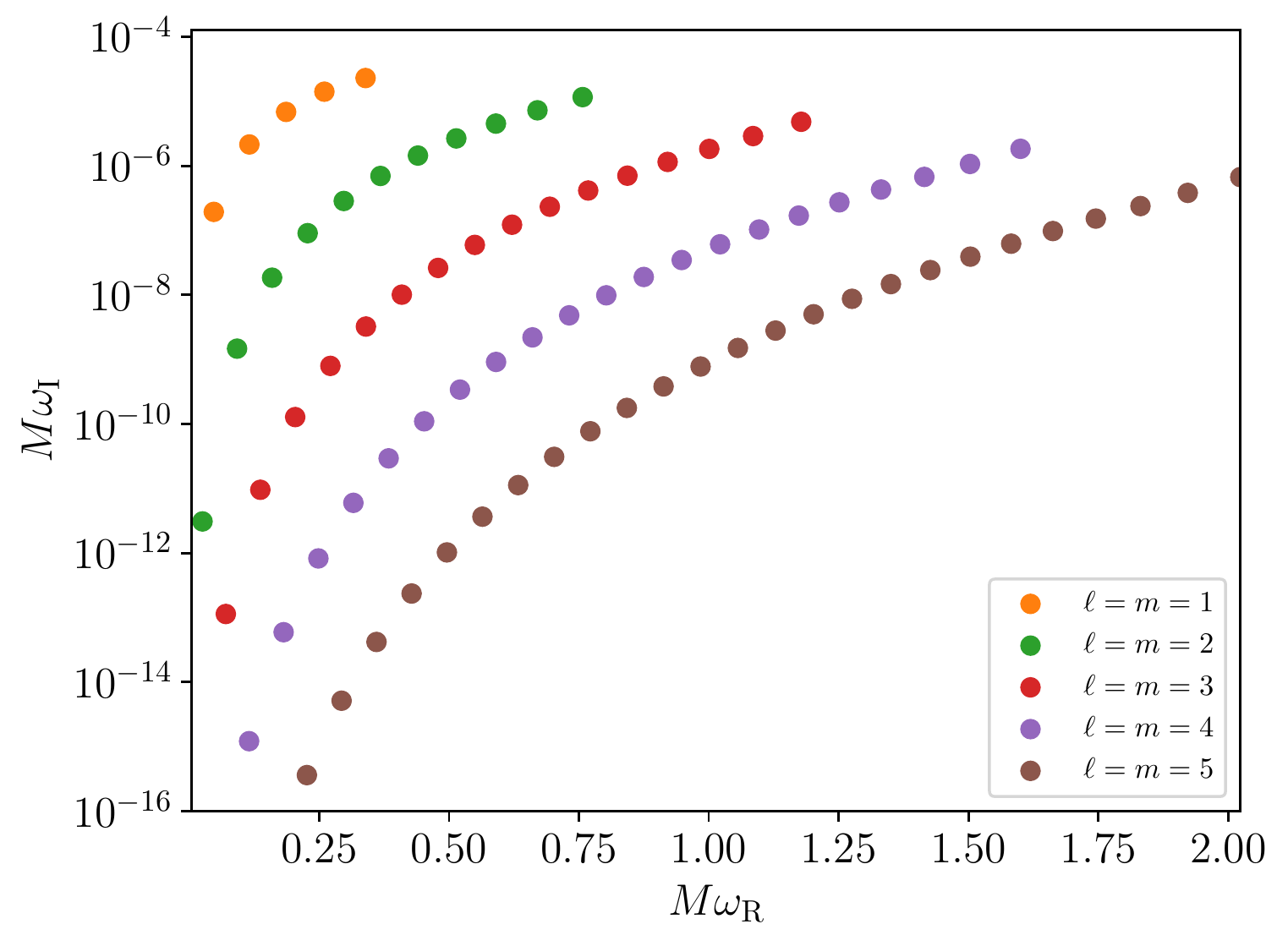}
\caption{QNMs of a spacetime with $a = 0.99$, $\epsilon = 10^{-3}$ for the six lowest multipoles with $\ell=m$. The figure includes both stable and unstable modes.
{\bf Top panel:} the dashed lines mark the superradiant condition $\omega_R=m\Omega$, which is the onset of the instability. Unstable modes are barely visible on this scale.
{\bf Bottom panel:} zoom of the top panel, showing only unstable modes with positive $\omega_R$. The next mode, with larger $\omega_R$ would be stable and would therefore fall below to the negative $\omega_I$ plane.
\label{fig:1000}}
\end{figure}
We have searched for the complex eigenfrequencies, which we write as
\be
\omega=\omega_R+i\omega_I\,,
\ee
for different spacetime spin parameter $a$ and surface location $r_0$ (or $\epsilon$).
We focus solely on the modes with $\ell=m$, for which there is an infinity of solutions, called overtones. The modes with $\ell=m$ have instability rates higher than the modes with $\ell>m$ (for details see Appendix~\ref{appendix:l_greather_m}), so we can proceed focusing only on $\ell=m$ modes, safe since we are concerned with the disappearance of instability. Although not the focus of this work, we also verified that in the spinless $a=0$ limit, large $\ell$ modes are extremely long-lived, as in Ref.~\cite{Cardoso:2014sna} (but in the latter work boundary conditions are imposed at the origin). In fact, we find that all modes are stable but their lifetime increases exponentially with $\ell$. This finding lends support to the claim that the geometry is a trapping geometry as we discussed in the introduction.

\begin{figure}[th]
  \centering
  \includegraphics[width=0.5\textwidth]{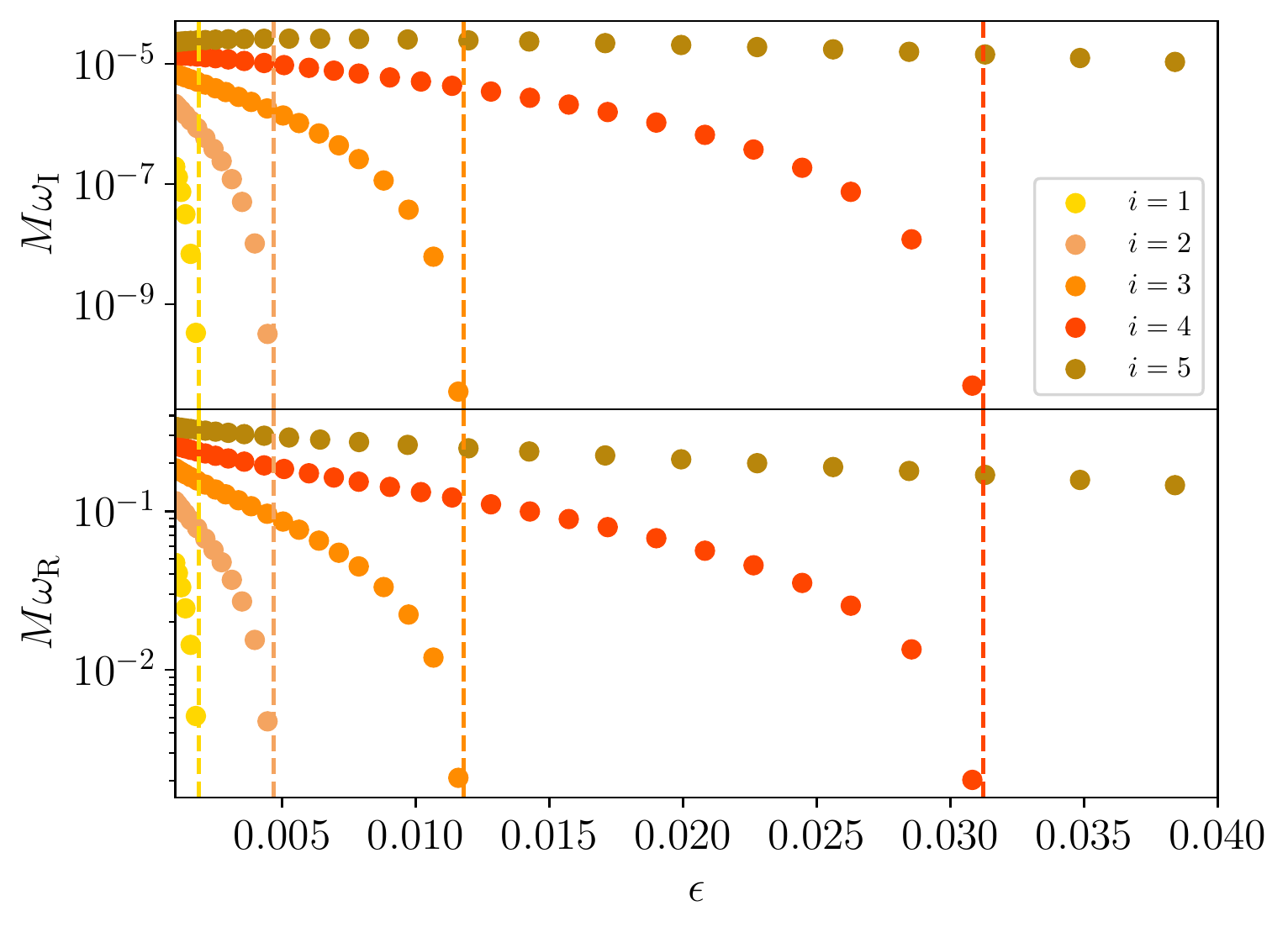}
  \caption{QNMs for $a=0.99 M$ and $\ell = m = 1$ as a function of $\epsilon$. We select five unstable modes at $\epsilon = 10^{-3}$ and follow them as $\epsilon$ increases. The figure shows that the modes eventually become zero-frequency modes, in this case at $\epsilon = 0.0019$, $0.0047$, $0.012$, $0.031$, $0.096$, marked by dashed lines. Our results indicate that all the unstable modes pass through zero frequency modes before becoming stable. \label{fig:epsilon_lm1}}
\end{figure}

Our results are summarized in Figs.~\ref{fig:1000}--\ref{fig:epsilon_lm1} and Table~\ref{table:modes}, and are always expressed in units of the spacetime mass, or equivalently, we set $M=1$. We first focus on $r_0$ close to $r_+$ ($\epsilon \ll 1$) so we recover previous results in the literature~\cite{Maggio:2017ivp,Maggio:2018ivz}. Fig.~\ref{fig:1000} shows a few tens of modes for the first six multipoles. There are a few aspects worth highlighting. The first is that there are both stable and unstable modes, and that the transition from stability to instability is well marked by the superradiant threshold $\omega_R=m\Omega$: modes for which $|\omega_R|<m\Omega$ are unstable, whereas the other modes are stable.
The dominant unstable mode is shown in Table~\ref{table:modes}.
Our results are consistent and in excellent agreement with the analytical and numerical results reported in Ref.~\cite{Maggio:2018ivz} for $\epsilon \ll 1$.

The threshold varies with $r_0$ (or $\epsilon$, cf. Eq.~\eqref{eq:surface}). Fig.~\ref{fig:epsilon_lm1} illustrates how the unstable modes converge to marginally stable modes 
when $\epsilon$ varies. These results, complementary to those in Refs.~\cite{Maggio:2017ivp,Maggio:2018ivz}, indicate that all unstable modes pass through zero frequency modes before they become stable~\cite{Maggio:2018ivz,Brito:2015oca}. When $\epsilon \ll 1$, our results are consistent with those of Refs.~\cite{Maggio:2017ivp,Maggio:2018ivz}.
\begin{figure}[th]
 \centering
 \includegraphics[width=0.5\textwidth]{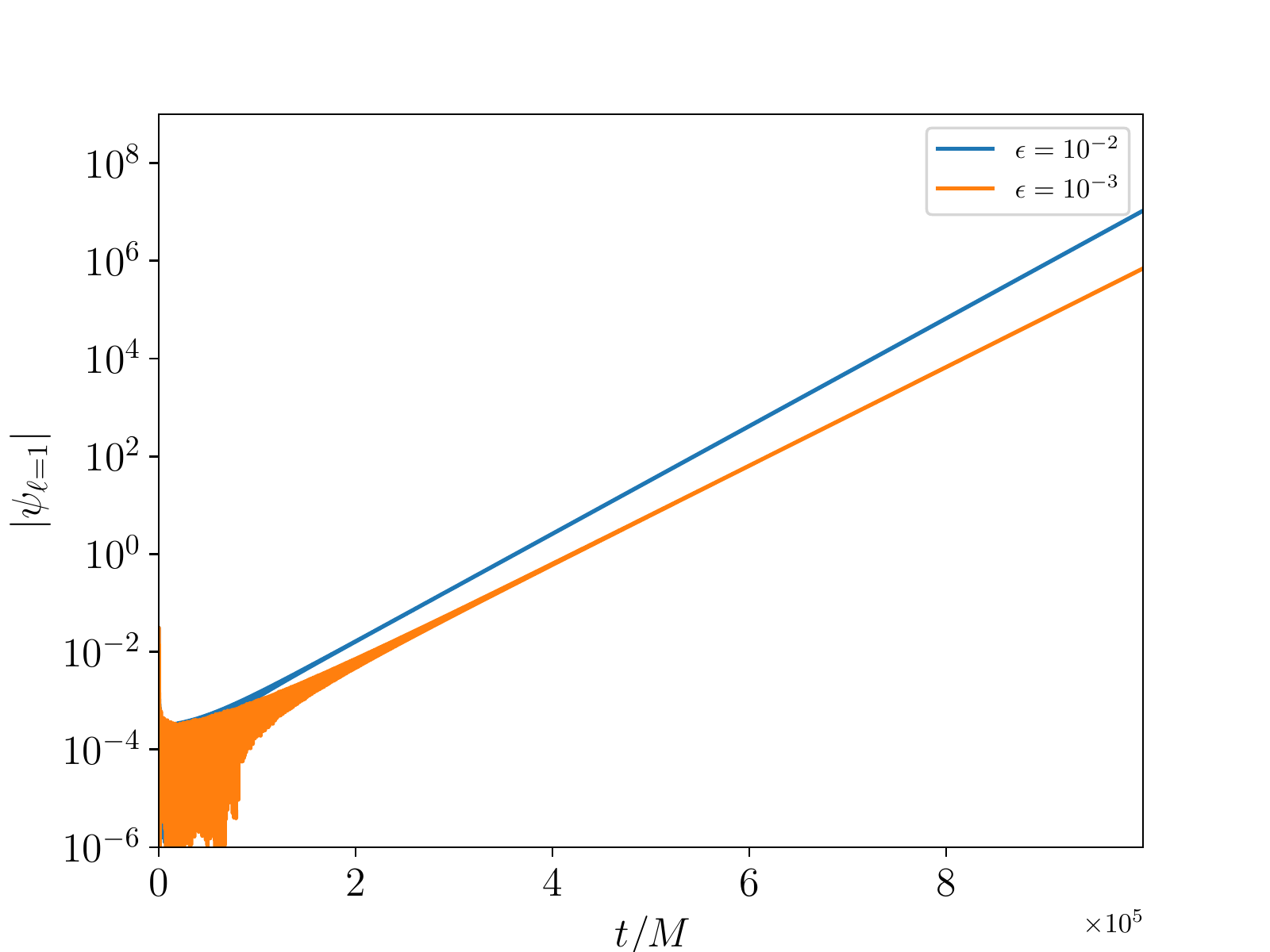}
 \caption{Time series data for the evolution of the field for $a=0.99M$ and $\ell=m=1$, up to $t=10^6 M$. 
The instability rate agrees well with frequency domain predictions. In particular, using a Prony method in the time-domain data we find a dominant unstable mode with $\omega \sim 0.339 + 2.32 \times 10^{-5}i$ for $\epsilon = 10^{-3}$, to be compared with the frequency-domain prediction in Table~\ref{table:modes}. The field is extracted at $r_* = -28.83,\, -47.25$ for $\epsilon=10^{-2},\,10^{-3}$, respectively. The instability details are independent on the extraction radii within numerical error.
\label{fig:a099_time_domain}}
\end{figure}
We complemented the frequency-domain analysis with the evolution in time of the wave equation subjected to Dirichlet boundary conditions \eqref{eq_bc_definition} at the surface of the object.
Figure~\ref{fig:a099_time_domain} shows the evolution for a rapidly spinning object with $a=0.99M$ and surface at $\epsilon=10^{-3},\,10^{-2}$. An initial transient is followed by a rapid exponential growth, with a growth rate which we extracted using Prony techniques. The results of the time-evolution are in very good agreement with the frequency-domain calculation of the dominant mode.

\subsection{The zero-frequency modes}\label{sec:zero_modes}
Our numerical study indicates that, at fixed overtone, the transition from stability to instability occurs via a zero-frequency mode. 
A similar feature had been observed numerically in analogue fluid geometries~\cite{Oliveira:2014oja} and used to explore analytically the transition from stability to instability~\cite{Hod:2017eld}. The existence of zero-frequency modes is -- to our knowledge -- far from trivial or obvious. Nevertheless, as we show below, they exist and we find simple analytical expressions requiring regularity at the boundaries. These modes provide a clean discriminator to understand better when our spacetime becomes linearly stable, so they merit a more detailed analysis (see also Ref.~\cite{Hod:2017cga}, where the investigation of zero-modes in this setup was initiated).
When $\omega=0$, ${}_0\mathcal{A}_{\ell m}(0)=\ell(\ell+1)$ and one finds a simple solution to the Klein-Gordon equation, in Boyer-Lindquist coordinates:
\beq
\Psi&=&a_1P_\ell^{-\frac{ima}{\Gamma}}\left(\frac{r-M}{\Gamma}\right)+a_2Q_\ell^{-\frac{ima}{\Gamma}}\left(\frac{r-M}{\Gamma}\right)\,,\nonumber\\
\Gamma&=&\sqrt{M^2-a^2}\,,
\eeq
where $P_\mu^{\nu}(z)$ and $Q_\mu^{\nu}(z)$ are generalized associated Legendre functions (type 2) of first kind and second kind~\cite{handmath}.
We will continue with units $M=1$ and focusing only on $\ell=m$ modes.
Requiring regularity at infinity, we find $a_1/a_2=-\pi/(2i)$. A zero-frequency mode appears when, and if, the surface at $r_0$ coincides with a zero of the function $\Psi$.
To analyze the zeros of $\Psi$, we define an auxiliary function
\be
\mathcal{F}(\alpha, m, x) = \frac{P_m^{-i \alpha m}\left(x\right)}{Q_m^{-i \alpha m}\left(x\right)} - \frac{2i}{\pi}\,.
\ee
where $m \in \mathbb{Z}^+$ and
\begin{align}
\alpha = \frac{a}{\sqrt{1 - a^2}}, \quad \alpha \in (0, +\infty)\,. \\
x = \frac{r - 1}{\sqrt{1 - a^2}}, \quad x \in (1, +\infty).
\end{align}
The function $\mathcal{F}(\alpha, m, x)$ has the same zeros as $\Psi$. Next, we analyze the zeros of the function analytically and numerically.
We will keep $\mathcal{F}$ as a function of $x(r)$ although we will impose the boundary condition at $r_0$.
%
For convenience of the analysis, we replace $m$ with a continuous variable $\mu$ in $\mathcal{F}(\alpha, \mu, x)$ where $\mu \in [1, +\infty)$ is a real number.
\begin{figure}[th]
  \centering
  \includegraphics[width=0.5\textwidth]{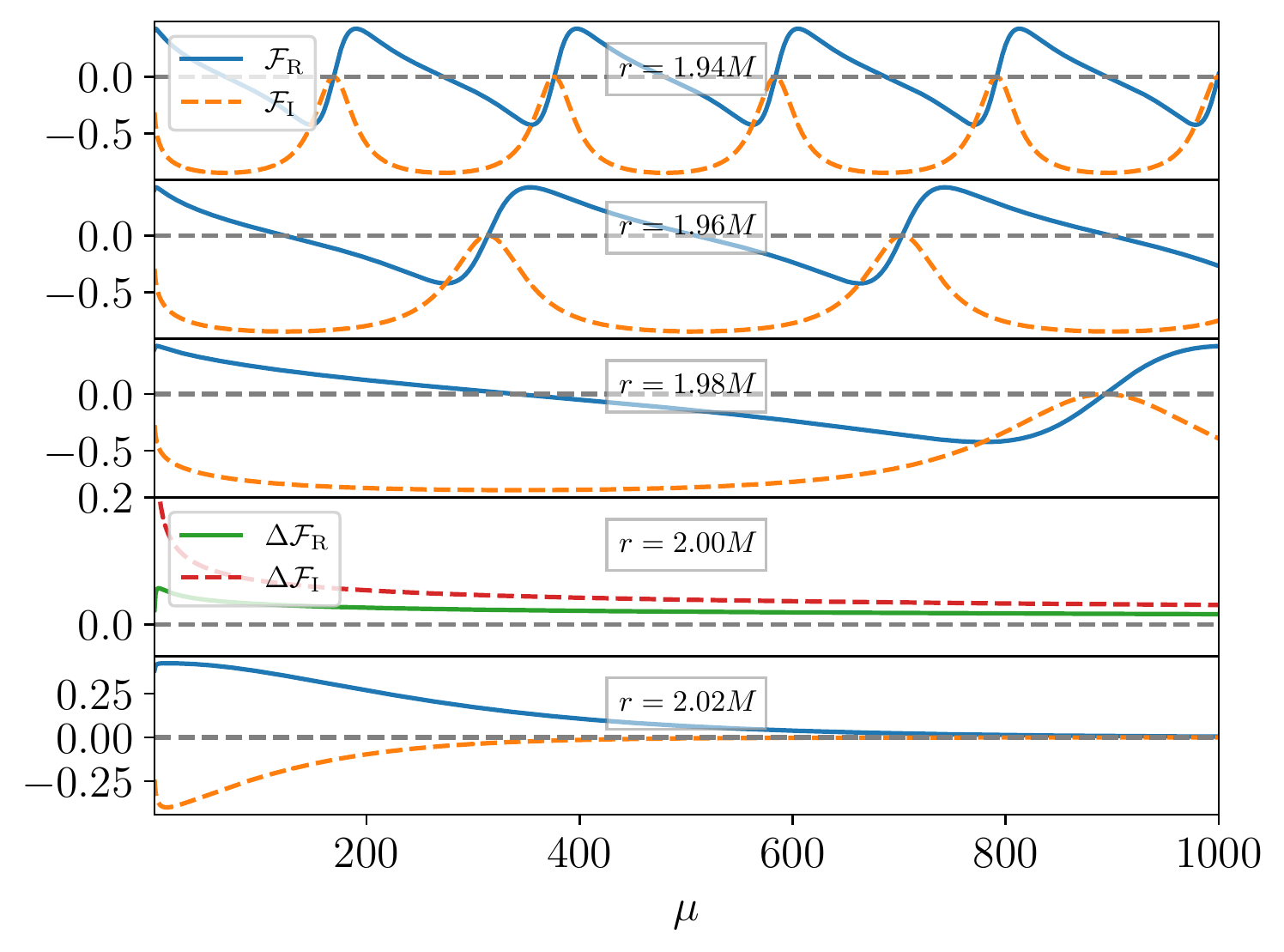}
  \caption{The function $\mathcal{F}(a, \mu, x)$ for $a = 0.99$ and fixed $r$ as a function of $\mu$, which is an asymptotically periodic function with period given by Eq.~\eqref{eq:period}. We define $\Delta \mathcal{F}_R = \Re{\mathcal{F} - \frac{2}{\sqrt{3}\pi}}$ and $\Delta \mathcal{F}_I = \Im{\mathcal{F} + i\frac{2}{\pi}}$. As $r$ gradually approaches $2M$, the zero points of $\mathcal{F}$ gradually disappear. For $r= 2M$, we derived analytically the result $\lim_{\mu\to+\infty} \mathcal{F}(a, \mu, r) = \frac{2}{\sqrt{3}\pi} - \frac{2i}{\pi}$, see discussion around \eqref{eq:asymptotic}. For $r> 2M$, we have $\lim_{\mu\to+\infty} \mathcal{F}(a, \mu, r) = 0$ (except in the neighbourhood of $r=2M$). Therefore, we conclude that there is no zero point for $r \geq 2M$.}
  \label{fig:period}
\end{figure}
Fig.~\ref{fig:period} shows how $\mathcal{F}(a, \mu, x)$ varies with $r$. We can see that the asymptotic behavior of the function depends on the surface location, namely it separates into three cases, $1 < x < \sqrt{1 + \alpha^2}$ ($r_+ < r < 2$) in the top three panels of Fig.~\ref{fig:period}, $x = \sqrt{1 + \alpha^2}$ ($r = 2$) in the fourth panel of Fig.~\ref{fig:period}, and $x > \sqrt{1 + \alpha^2}$ ($r > 2$) in the last panel of Fig.~\ref{fig:period}. 
We deal with these cases separately and we use the asymptotic expansion of $\mathcal{F}(\alpha, \mu, x)$ for $\mu \to +\infty$ given in Ref.~\cite{https://doi.org/10.48550/arxiv.1609.08365}\footnote{Note that the definition of the generalized associated Legendre function in Ref.~\cite{https://doi.org/10.48550/arxiv.1609.08365} is different from ours, as we take into account when expressing asymptotic properties.}.

When $1 < x < \sqrt{1 + \alpha^2}$ (corresponding to a surface within the ergoregion, $r_+ < r < 2$), we can see that $\mathcal{F}(a, \mu, x)$ is an asymptotically periodic function of $\mu$ at fixed $a$ and $x$. In particular, at large $\mu$
\begin{align}
\mathcal{F} &\sim -\frac{2 i}{\pi -2 \pi f_1 f_2 f_3 f_4 f_5} - \frac{2i}{\pi}\,,\\
f_1 &= \left(\alpha ^2+1\right)^{-i \alpha  \lambda +\lambda -\frac{1}{2}}\,,\\
f_2 &= \left(x^2-1\right)^{-i \alpha  \lambda}\,, \\
f_3 &= \left(x+i f_6\right)^{-2 \lambda }\,, \\
f_4 &= f_6+i x\,, \\
f_5 &= \left(f_6+\alpha  x\right)^{2 i \alpha  \lambda }\,,\\
f_6 &= \sqrt{\alpha ^2-x^2+1}\,,
\end{align}
and has infinite zeros
\be
\lambda = \frac{\tan^{-1}(\frac{x}{f_6})}{\log p(\alpha, x)} + n \mathcal{P}(\alpha, x)\label{eq:zeros}\,,
\ee
where $n \in \mathbb{Z}$,
\be
p(\alpha, x) = \left(\frac{\left(\alpha ^2+1\right) \left(x^2-1\right)}{\left(f_6+\alpha  x\right)^2}\right)^{\alpha} e^{2 \tan^{-1}\frac{f_6}{x}} \,,
\ee
and its period $\mathcal{P}(\alpha, x)$ is
\be
\mathcal{P}(\alpha, x) = -\frac{2 \pi }{\log p(\alpha, x)}\,.\label{eq:period}
\ee
It is easy to verify that $\pdv{\mathcal{P}}{x} > 0$, $p > 0$ and $\pdv{p}{x} > 0$, so the period $\mathcal{P}$ increases as $x$ increases, consistently with Fig.~\ref{fig:period}.

For $x=1$, we have $\mathcal{P}(\alpha, 1) = 0$. Thus, for surfaces placed close to the horizon at $r_+$ there are a large number of zeros, and hence a large number of unstable modes.
For $x=1+\delta$ and $\delta \ll 1$, we obtain the asymptotic expansion of $\mathcal{F}(a, \mu, x)$ near the horizon. If we keep only the leading term in $\delta$, then we find a periodic function of $\log(\delta)$ with period $2 \pi / (\alpha m)$ and zero points at
\beq
&&\log\delta + \frac{2\pi}{\alpha m}n \nonumber\\
&=& \frac{i}{am}\log \left(\frac{2^{-i \alpha  m } \Gamma (1-i \alpha  m ) \Gamma (i \alpha  m +m +1)}{\Gamma (1 + i \alpha  m) \Gamma (-i \alpha  m +m +1)}\right)\,,\nonumber
\eeq
where $n \in \mathbb{Z}$ and $\delta \to 0^+$ corresponds to $n \to +\infty$. Thus, as we bring the Dirichlet condition closer to the horizon, there are more zero frequency modes, i.e. we can get an infinite number of superradiant modes in this way. 
This result is consistent with the findings in Ref.~\cite{Hod:2017cga} and Ref.~\cite{Maggio:2018ivz} (c.f. Eq.~(29) with $q=2n$ where $n=1,2,...$).

\begin{figure}[th]
  \centering
  \includegraphics[width=0.5\textwidth]{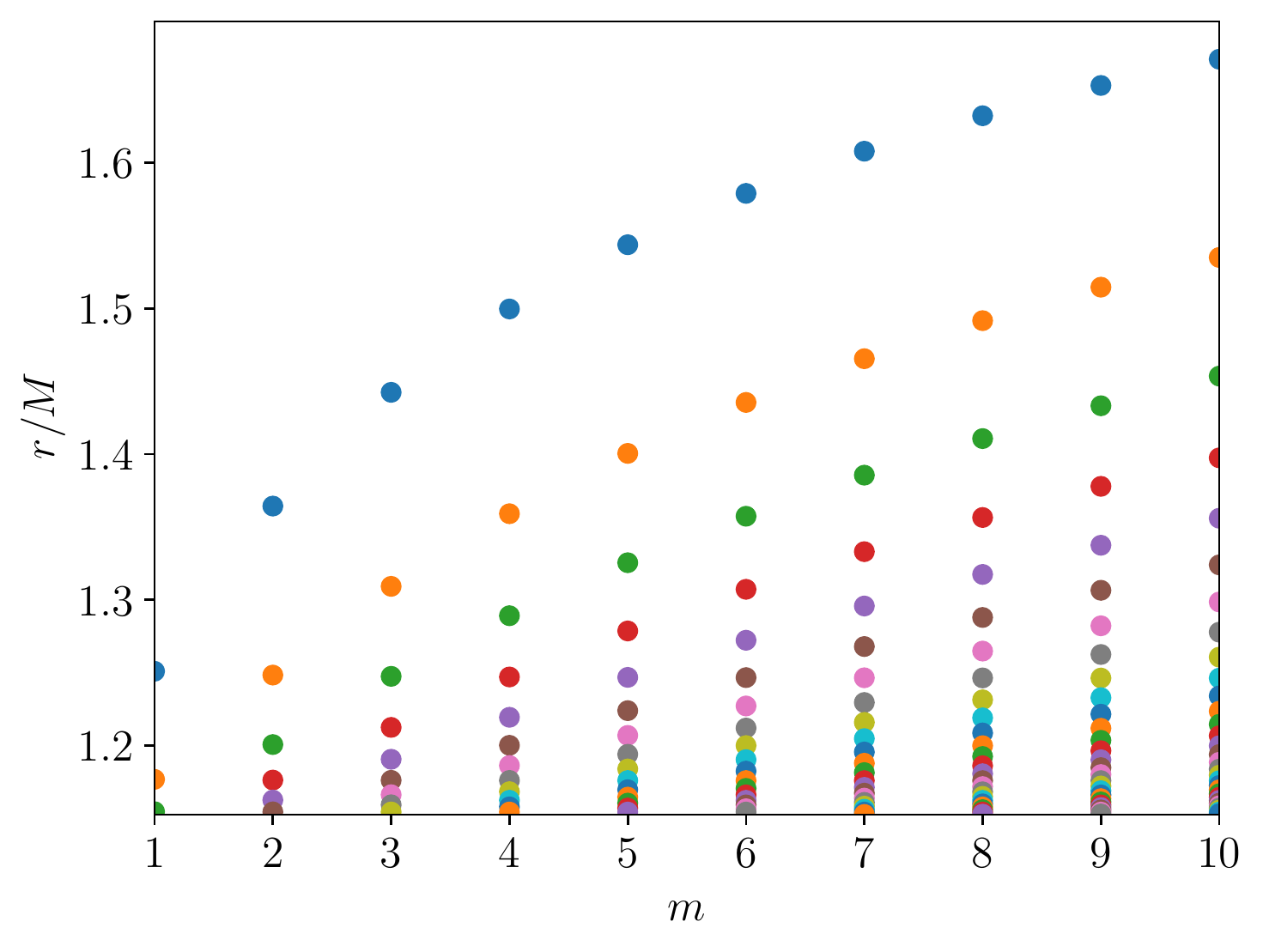}
  \caption{The surface location of zero-frequency modes with $a=0.99M$ as a function of $\ell = m$. As discussed in the main text, the outermost zero mode is always at $r<2M$ in the limit $m \to \infty$. If we consider $m \in \mathbb{Z}^+$ as $\mu \in [1, +\infty)$, then each zero-mode family is a continuous line as function of $\mu$. We use different colors to distinguish the zero points connected by different lines.}
  \label{fig:zeros}
\end{figure}
\begin{figure}[th]
  \centering
  \includegraphics[width=0.5\textwidth]{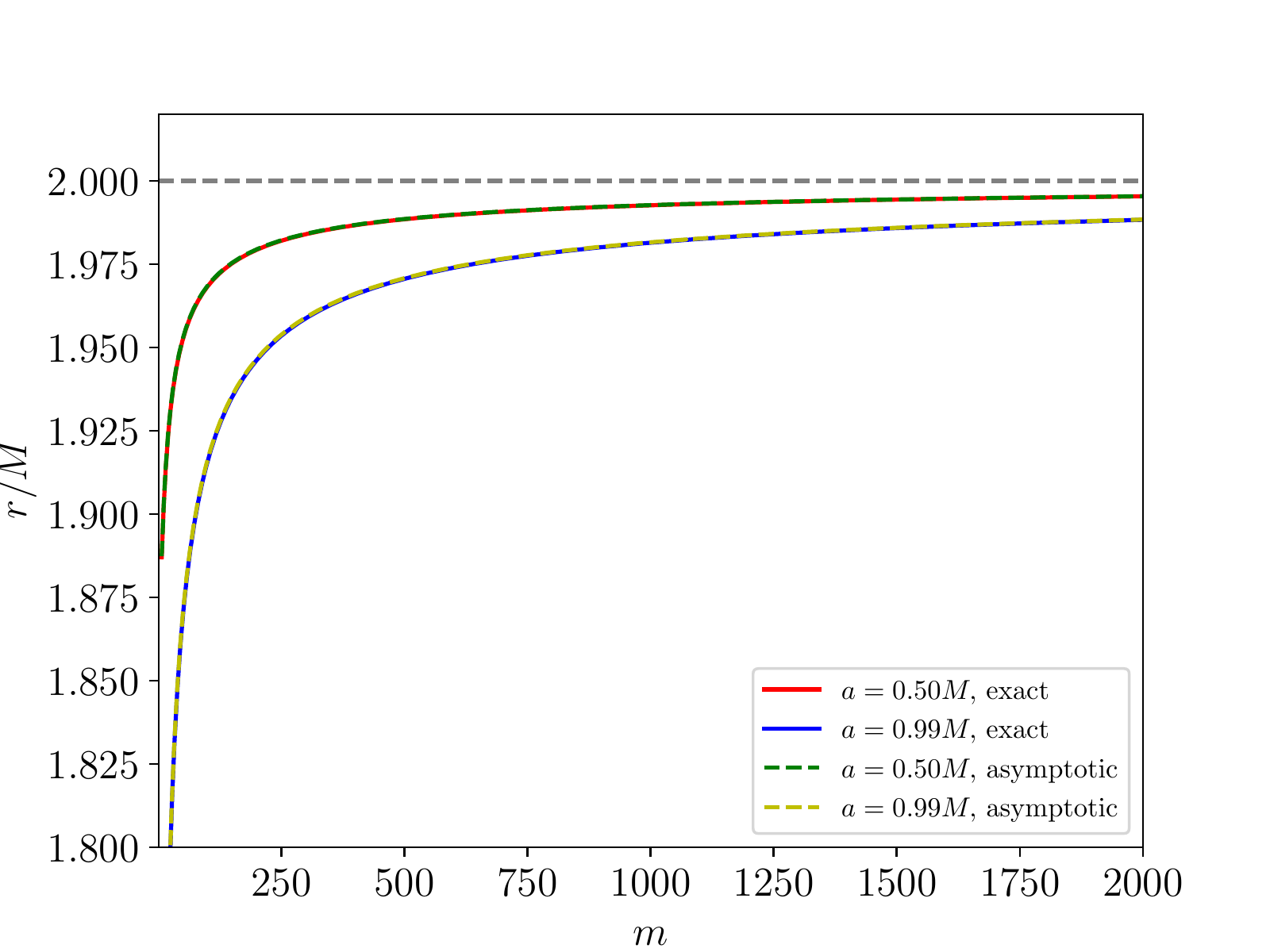}
  \caption{Outermost zero-frequency modes as a function of $\ell = m$ for two representative values of the spin, 
  $a/M=0.50,0.99$. A polynomial fit indicates that the critical surface lies at $r_{\rm crit}=1.998M, \,1.999M$ 
	at large $m$ for $a/M=0.99, 0.5$ respectively, compatible with the location of the equatorial ergosurface. The dashed lines are obatined by using the inverse function of Eq.~\eqref{eq:zeros} with $n = 1$.}
  \label{fig:outermost-zeros}
\end{figure}
On the other hand, for $x=\sqrt{1+\alpha^2}$, we have $p(\alpha, \sqrt{1 + \alpha^2}) = 1$, then we can get $\mathcal{P} \to +\infty$ for $p \to 1^-$. 
Thus, the number of zero modes vanishes asymptotically when the surface approaches the equatorial ergoregion boundary at $r_0=2M$.
This zero-frequency behavior is shown in Fig.~\ref{fig:period}.

In summary, our analytical results show that there exist zero modes for objects with ergoregions -- as had also been previously discussed~\cite{Hod:2017cga} -- and their existence is {\it precisely} delimited by the equatorial ergoregion. An example of this behavior is shown in Figs.~\ref{fig:zeros},~\ref{fig:outermost-zeros}. Figure~\ref{fig:zeros} shows the surface location of the various zero-modes, for different $\ell=m$ modes. As discussed before, at fixed $m$ there are several solutions sustaining zero modes. The outermost solution (i.e., the largest $r_0$, blue dots in Fig.~\ref{fig:zeros}) are shown in Fig.~\ref{fig:outermost-zeros} but now for the first $2000$ multipoles. We also show the corresponding modes for $a=0.5M$.
In other words, these results strongly suggest that as long as the surface is placed within the ergoregion, there will be zero-modes and hence linear instabilities.
Thus, our findings are consistent with the generic proofs that asymptotically flat, horizonless spacetimes with ergoregions are unstable~\cite{Friedman:1978wla,1978CMaPh..63..243F,Moschidis:2016zjy,Vicente:2018mxl,Brito:2015oca}.

When the surface location is at the outermost boundary of the ergosphere ($x = \sqrt{1+\alpha^2}$ or $r = 2M$), then two saddle points coalescence (i.e. Eq.~(4.3) and Eq.~(4.4) in Ref.~\cite{https://doi.org/10.48550/arxiv.1609.08365}). Then,
\begin{equation}
\lim_{\mu\to+\infty} \mathcal{F}(a, \mu, \sqrt{1 + \alpha^2}) = \frac{2}{\sqrt{3}\pi} - \frac{2i}{\pi}\,,\label{eq:asymptotic}
\end{equation}
a limit which is well captured by our numerics at $a=0.99$ (cf. panel 4 in Fig.~\ref{fig:period}).

The third case where the radius is greater than the outermost boundary of the outer ergosphere ($x > \sqrt{1+\alpha^2}$ or $r > 2$) (except in the neighbourhood of the $\sqrt{1+\alpha^2}$), we have
\begin{equation}
  \lim_{\mu\to+\infty} \mathcal{F}(a, \mu, x) = 0 \,.
\end{equation}
Combining Fig.~\ref{fig:period}, we conclude that there are no zero-frequency modes when $x \ge \sqrt{1 + \alpha^2}$ ($r \ge 2$). Especially, $x = \sqrt{1 + \alpha^2}$ ($r = 2$) is not a zero point even in the limit $m \to \infty$. This indicates that all superradiant modes disappear before reaching the outermost boundary of the outer ergosphere. Note that when restricting $\mu \in [1, +\infty)$ to $m \in \mathbb{Z}^+$, the zeros of $\mathcal{F}(a, m, x)$ may even disappear earlier when $x \to \sqrt{1 + \alpha^2}$.
%

\subsection{Is there a new family of modes?}\label{sec:newmodes}
\begin{figure}[th]
  \centering
  \includegraphics[width=0.5\textwidth]{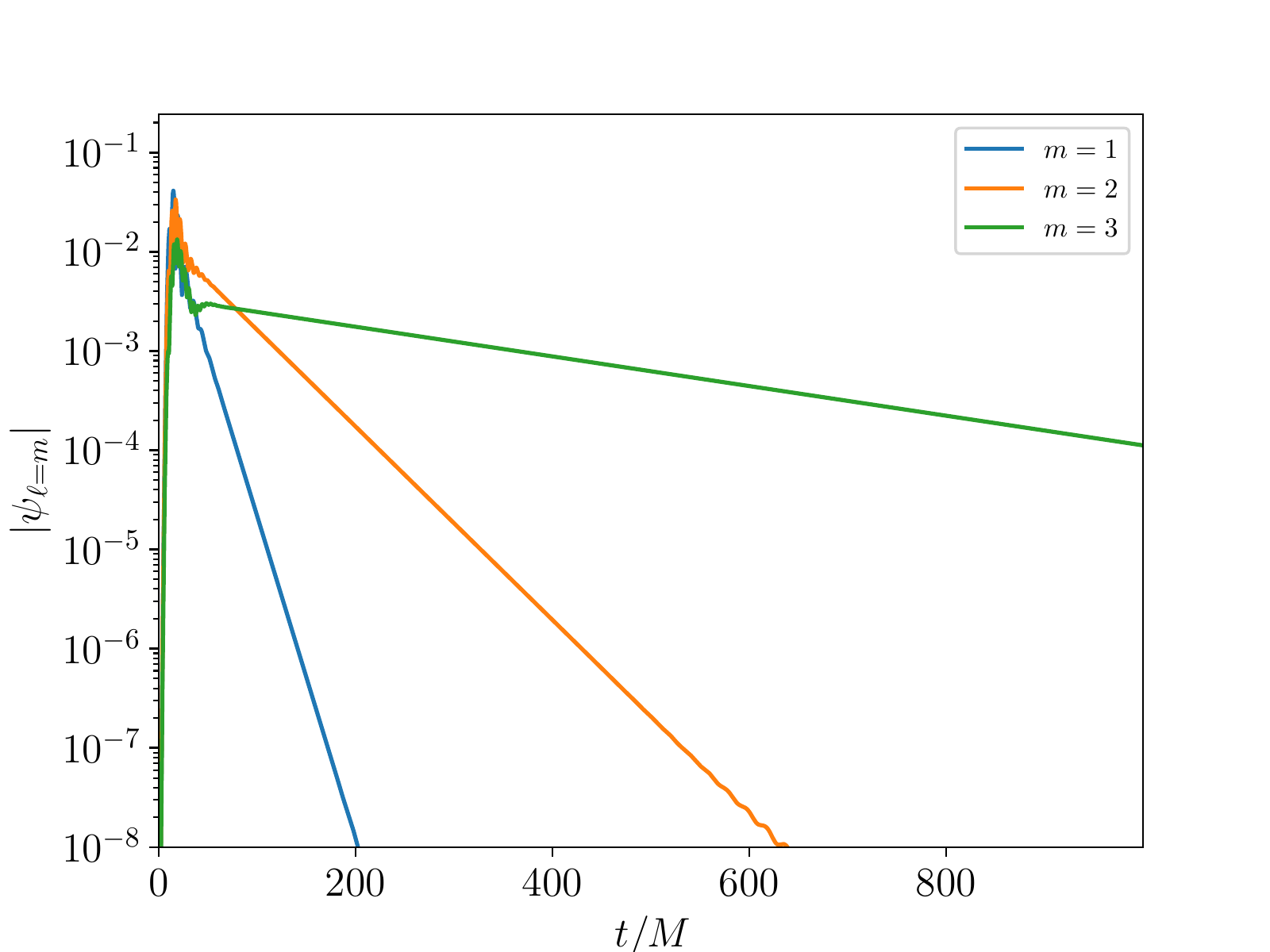}
	\includegraphics[width=0.5\textwidth]{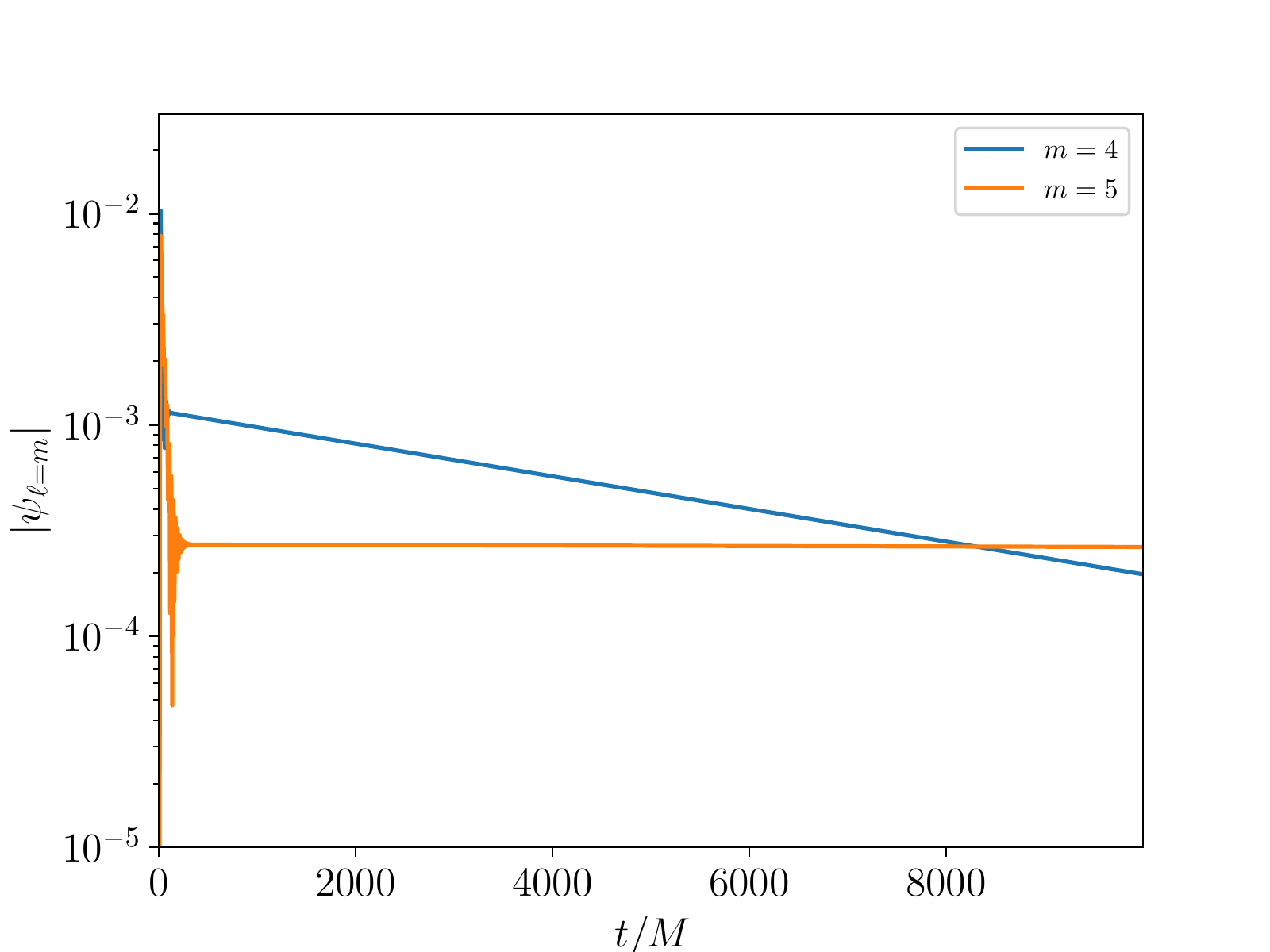}
\caption{Time series data for the evolution of a scalar field in the geometry of an object with $a=0.5M$ and a surface at $r_0=2M$. The field is extracted at $r_*=-3.69$, but the overall behavior is independent of the extraction radii.
{\bf Top Panel:} Evolution of the multipoles $m = 1, 2, 3$, up to $t=10^3 M$. Using a Prony method, we estimate a dominant mode with $M\omega=-0.338-7.48 \times 10^{-2}i$, $-0.444-2.24\times 10^{-2}i$ and $-0.532-3.44\times 10^{-3}i$ for $m =1,2, 3$ respectively. These estimates compare well with the frequency-domain data in Table~\ref{table:modes_2M}.
{\bf Bottom Panel:} Same as top panel, for multipoles $m=4,5$ which are longer-lived, up to $t = 10^4 M$. 
A Prony method now yields $M\omega \sim -0.600 - 1.78 \times 10^{-4}i$ and $-0.652-2.65\times 10^{-6}i$ for $m = 4,5$, which are still in good agreement with frequency-domain predictions, despite the rather large timescales now involved. \label{fig:a050_2M_time_domain_m123}}
\end{figure}
%
%
Our results are a strong indication that zero-frequency modes do not exist when $r_0>2M$. Thus, the ergoregion family of modes
cannot be the explanation for the findings of Ref.~\cite{Cunha:2022gde}. However, we are still left with the possibility that a linearized instability exists, not associated with any zero-frequency mode. In particular, it is possible that a new instability mechanism sets in for spinning spacetimes without ergoregions but with light rings.
To investigate this possibility, we did a thorough search of unstable modes in the spectrum of the problem when $r_0\geq 2M$ and $a=0.5M$ (as discussed above, for $a<\sqrt{2}M/2$ the unstable equatorial light ring lies outside the ergoregion). We found no unstable mode. The evolution of initial data, using time-domain methods, also shows no hints of a physical instability, but it does show clearly the dominance of long-lived modes. Typical examples are shown in Fig.~\ref{fig:a050_2M_time_domain_m123}.

\section{Discussion}
Our results establish that there are exponentially growing modes in a horizonless Kerr geometry, when it contains ergoregions. The instability is connected continuously to zero-frequency modes which cease to exist for surface locations outside the ergoregion, i.e., for $r_0\geq 2M$. An important point for the context at hand is that the ergoregion instability has timescales $\tau \gtrsim 10^5M$~\footnote{For $a/M = 0.99$ and $m = 1$, the maximum instability occurs at $\epsilon \sim 0.0056$, for $a/M = 0.99$ and $m = 2$ at $\epsilon \sim 0.0052$, and the relevant timescales are of order $\sim 10^6M$~\cite{Maggio:2018ivz}.}.
One of our main motivations was to understand possible new linear mechanisms in the absence of an ergoregion but when light rings are present, possibly explaining the findings of Ref.~\cite{Cunha:2022gde} and the relatively short timescales reported in that work (possibly too short for a nonlinear mechanism).
We found no evidence of new instabilities on timescales $\lesssim 10^{5}M$.

\begin{acknowledgments}
Z.Z.\ acknowledges financial support from China Scholarship Council (No.~202106040037).
V.C.\ is a Villum Investigator and a DNRF Chair, supported by VILLUM FONDEN (grant no.~37766) and by the Danish Research Foundation.  V.C.\ acknowledges financial support provided under the European
Union's H2020 ERC Advanced Grant ``Black holes: gravitational engines of discovery'' grant agreement
no.\ Gravitas–101052587. Views and opinions expressed are however those of the author only and do not necessarily reflect those of the European Union or the European Research Council. Neither the European Union nor the granting authority can be held responsible for them.
E.M. acknowledges funding from the Deutsche Forschungsgemeinschaft (DFG) - project number: 386119226.
This project has received funding from the European Union's Horizon 2020 research and innovation programme under the Marie Sklodowska-Curie grant agreement No 101007855.
We acknowledge financial support provided by FCT/Portugal through grants 
2022.01324.PTDC, PTDC/FIS-AST/7002/2020, UIDB/00099/2020 and UIDB/04459/2020.
\end{acknowledgments}
%
%
%
%
%
%
%
%

\appendix
\section{Mimicking the ergoregion instability with a two-dimensional toy model}\label{appendix:interior}
\begin{figure}[th]
  \centering
  \includegraphics[width=0.5\textwidth]{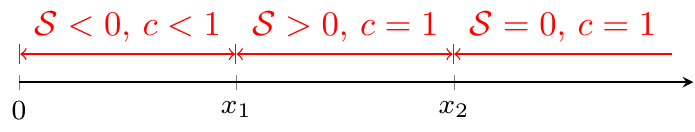}
  \caption{Two-dimensional star model with different sound speed, $c$, and Lorentz-violating factor, $\mathcal{S}$, in different intervals. $x < x_1$ is the interior of the star, $x_1 < x < x_2$ is the ergoregion mimicker, and $x > x_2$ is the vacuum case. \label{fig:1d_axis}}
\end{figure}
Here, we discuss a simple two-dimensional toy model, where we implement superradiance via a Lorentz-violating term in the Klein-Gordon equation (following the original work by Zel'dovich~\cite{zeldovich1,zeldovich2}) and we mimic the star interior assigning a sound speed different from unity in its interior. To be specific, we consider the second order partial differential equation
\begin{equation}
\left(- \frac{1}{c^2}\pdv[2]{}{t} + \pdv[2]{}{x} + \mathcal{S}\pdv{}{t}\right) \Phi(t, x) = 0\,,
\end{equation}
in the half-line $x>0$, where $x=0$ stands for the star center. The speed of propagation is $c < 1$ in the interior of the star $x<x_1$, and we assume that the star is an absorber, with $\mathcal{S}<0$. Between $x_1<x<x_2$ there is an ergoregion, which we model by adding a Lorentz-violating parameter $\mathcal{S}>0$. For $x>x_2$ we have vacuum and $c=1,\mathcal{S}=0$.
By assuming an harmonic ansatz,
\begin{equation}
\Phi(t, x)=\chi(x)e^{-i \omega t}\,,
\end{equation}
we can get
\begin{equation}
\chi''(x) + \left(\frac{\omega^2}{c^2} - i \mathcal{S} \omega\right) \chi(x) = 0\,.
\end{equation}
The generic solution is
\begin{equation}
\begin{aligned}
\chi =& c_1^{(i)} \exp(\frac{(-1)^{\frac{1}{4}} x \sqrt{i\omega + c^2 \mathcal{S}}\sqrt{\omega}}{c}) \\
&+ c_2^{(i)} \exp(-\frac{(-1)^{\frac{1}{4}} x \sqrt{i\omega + c^2 \mathcal{S}}\sqrt{\omega}}{c}) \,,
\end{aligned}
\end{equation}
where $c_1^{(i)}$ and $c_2^{(i)}$ are constants in $i$-th interval, which is determined by the boundary condition $\chi(0) = 0$, the outgoing boundary condition $\chi(x > x_2) \sim \exp(i \omega x)$ and connection conditions ($\chi(x)$ and $\chi'(x)$ is continuous) at $x_1$ and $x_2$.

\begin{figure}[th]
  \includegraphics[width=0.5\textwidth]{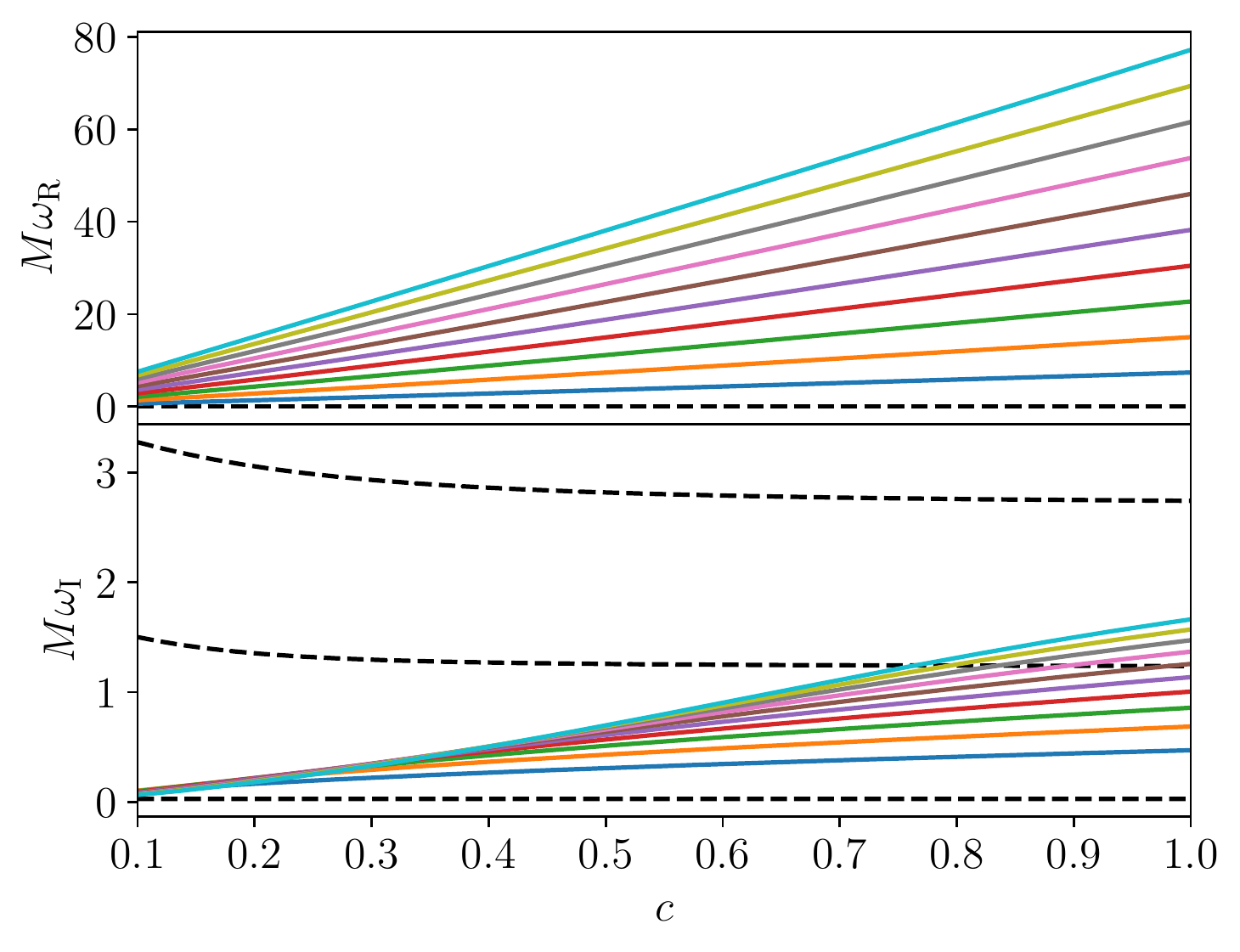}\\
  \includegraphics[width=0.5\textwidth]{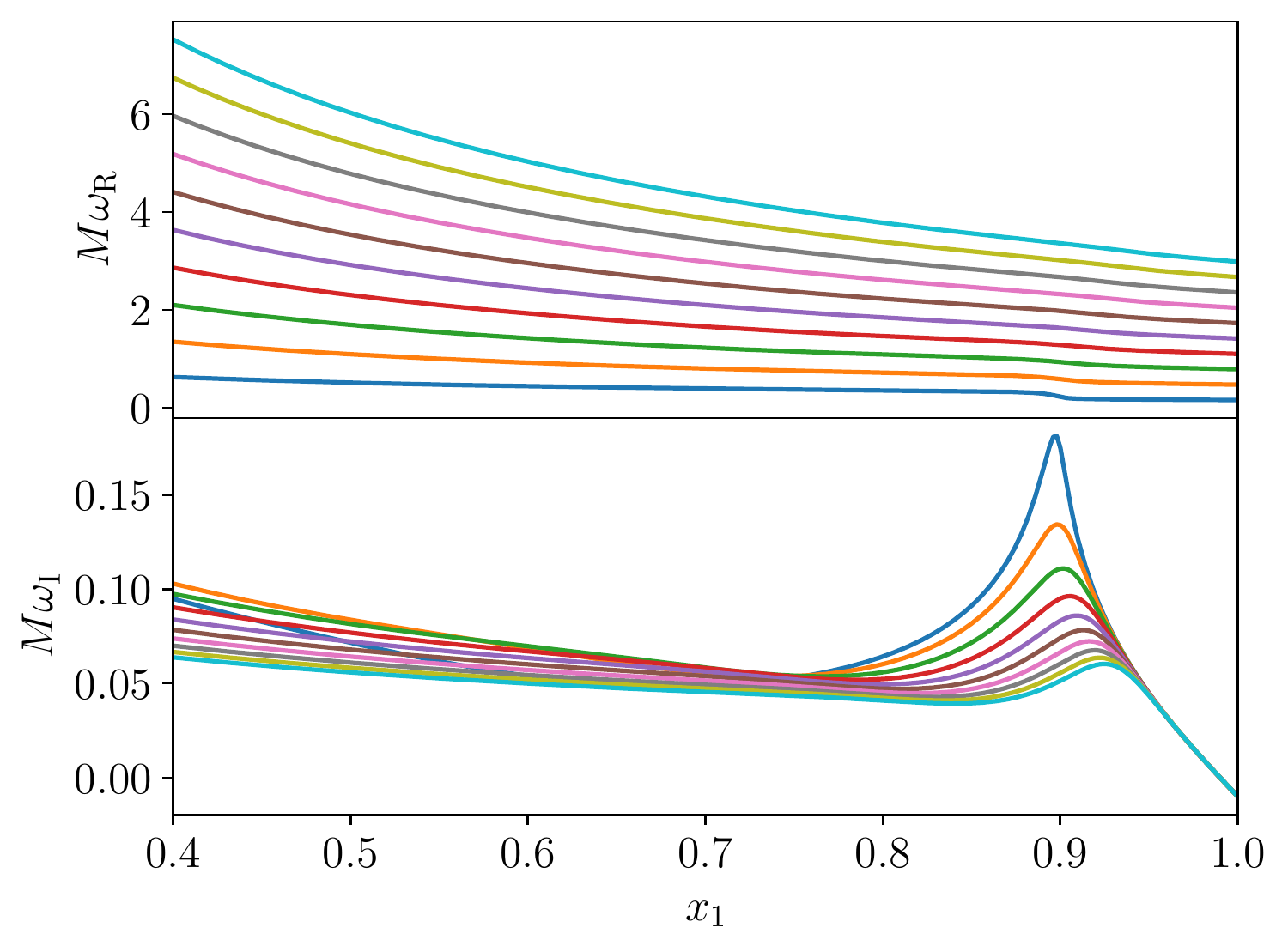}
  \caption{QNMs of the two-dimensional model with $x_2 = 1$ and $\mathcal{S} = 100$ for $x_1<x<x_2$. The star is not absorbing, so we set $\mathcal{S}=0$ in its interior $x<x_1$. We show the 10 most unstable modes, labeled by different colors.
  {\bf Top panel:} QNM frequencies as a function of the sound speed $c$ with $x_1 = 0.4$. The black dashed lines are purely imaginary QNMs.
  {\bf Bottom panel:} QNM frequencies as a function of the surface location $x_1$ when $c = 0.1$. Notice that when $x_1$ approaches $x_2$, the instability vanishes. Although not included in this bottom panel, the instability rate of the purely imaginary modes also vanishes when $x_1$ approaches $x_2$.
  \label{fig:2d}}
\end{figure}
\begin{figure}[th]
  \includegraphics[width=0.5\textwidth]{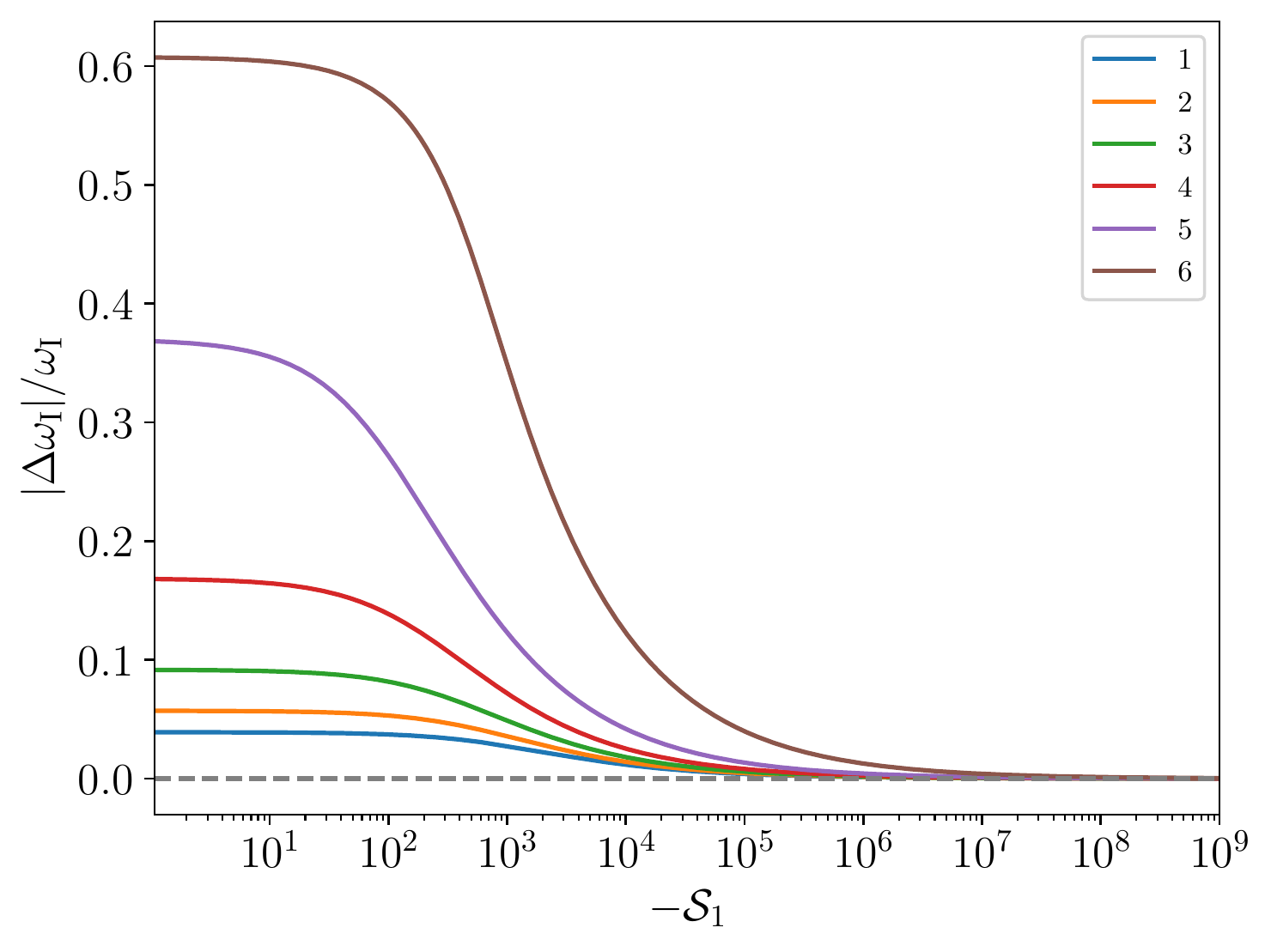}
  \caption{QNMs of the two-dimensional model with $x_1 = 0.4$, $x_2 = 1$ and $\mathcal{S} = 100$ for $x_1<x<x_2$. We vary $\mathcal{S} = \mathcal{S}_1$ in the star interior $x<x_1$. These modes are pure imaginary modes, and the imaginary parts are $9.50$, $6.12$, $3.61$, $1.81$, $0.64$ and $0.07$ in the limit $\mathcal{S}_1 \to -\infty$ for labels $1$ to $6$, respectively. In this limit, the modes converge to modes obtained by imposing Dirichlet boundary conditions at $x_1$, $\chi(x_1) = 0$.
  \label{fig:2d_S1}}
\end{figure}
Results are summarized in Fig.~\ref{fig:2d}, where for now we let $\mathcal{S}=0$ in the star interior, i.e., the star is not absorbing. The top panel shows the dependence of the characteristic ringing frequency and the instability timescale on the speed of waves inside the star. We show the first 10 most unstable modes, labeled by different colors. The star material in this model is slowing the back-and-forth process of negative-energy waves, thus delaying the growth. This explains why the frequency decreases when $c$ decreases inside the star, and also why $M \omega_I$ decreases too. The only exception concerns purely imaginary modes, the instability rate of which remains roughly constant when the sound speed changes. These are modes which damp out very quickly inside the star and therefore are not affected by sound speed. In fact, and because of this property, we will see below that such modes reduce to modes calculated with Dirichlet conditions at the surface. There is no new feature appearing when the sound speed varies, the structure of the modes only changes at the qualitative level. The bottom panel on the other hand, shows the dependence of the instability on the location of the surface of the star. There are two noteworthy features here:
when $x_1$ -- the location of the surface -- varies, the main features of the instability remain, lending support to our procedure of simply imposing Dirichlet conditions at the surface.
The second noteworthy feature is that when $x_1\to x_2$ the instability disappears, since there is no ergoregion anymore. 

Let us now turn the absorption on, in the star interior. In parallel with conducting materials in electromagnetism, one could infer that dealing with the star interior is equivalent to simply imposing Dirichlet conditions at its surface. This is also the underlying rationale behind the simple model in the main text. To understand if this is true, we calculate the characteristic frequencies imposing boundary conditions at $x=0$, and compare them with the spacetime with no star, where Dirichlet conditions are imposed instead at the surface, $x=x_1$. Results are summarized in Fig.~\ref{fig:2d_S1}. The figure shows the relative difference in $\omega_I$ when calculating the modes imposing boundary conditions at the center of the star ($x=0$) or at its surface ($x=x_1$) for a very absorbing material, as a function of the Lorentz-violating factor in the star interior. The figure shows, in the first place, that no new qualitative feature arises when imposing conditions at the star surface. But it also shows that, in the limit where absorption is very large, $\mathcal{S}_1 \to -\infty$, the QNM frequencies of the two problems {\it are the same}. This justifies well our usage of the exterior Kerr spacetime in the main body of this work, with Dirichlet conditions at its surface.

\section{$\ell > m$ cases}\label{appendix:l_greather_m}

\begin{figure}[th]
  \includegraphics[width=0.5\textwidth]{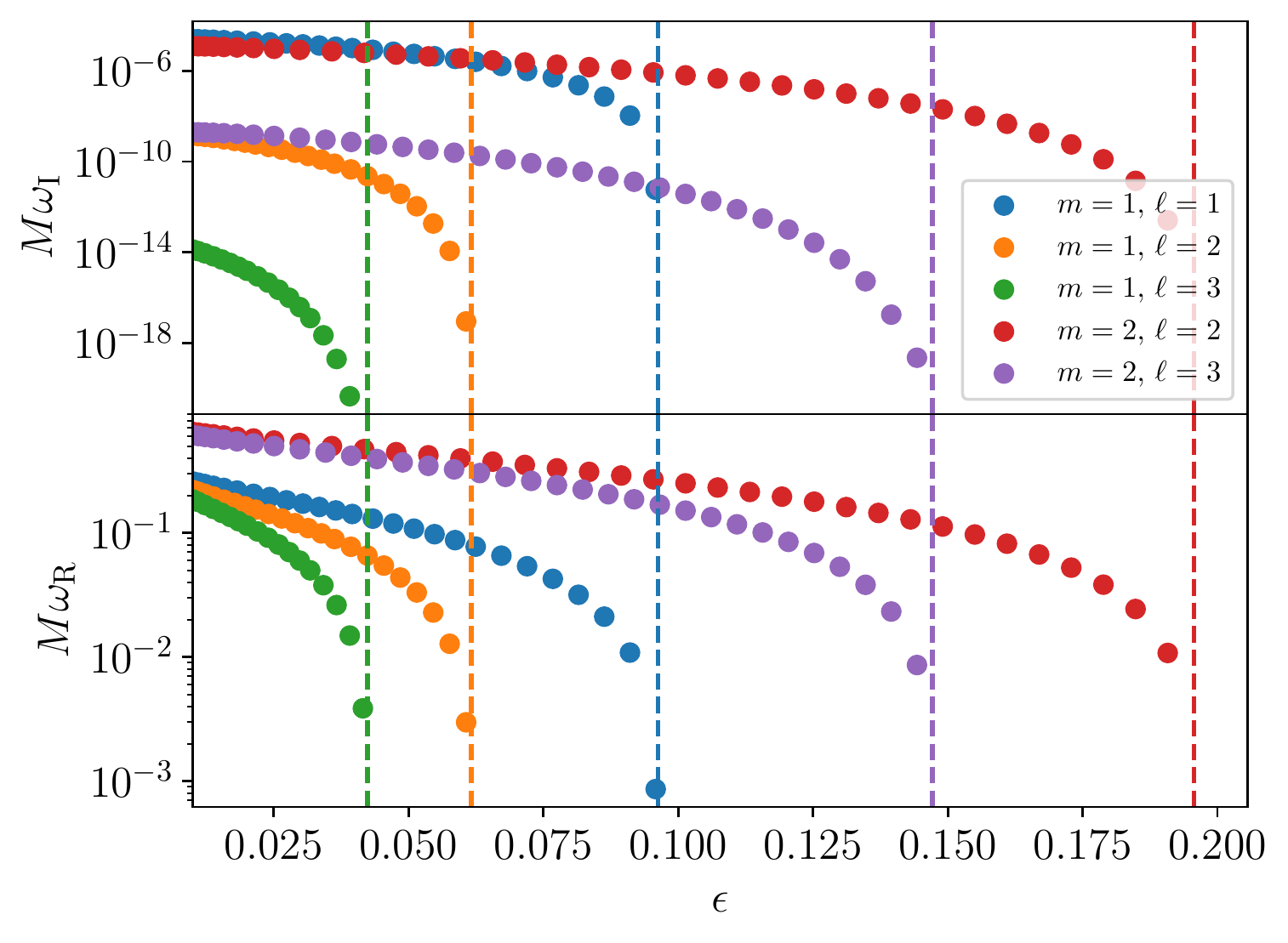}
  \caption{Dominant QNMs for $a = 0.99M$ as a function of $\epsilon$. The figure shows that the modes eventually become zero-frequency modes, in this case at $\epsilon = 0.096$, $0.062$, $0.042$, $0.20$, $0.15$, marked by dashed lines. \label{fig:qnm_lm}}
\end{figure}

\begin{figure}[th]
  \includegraphics[width=0.5\textwidth]{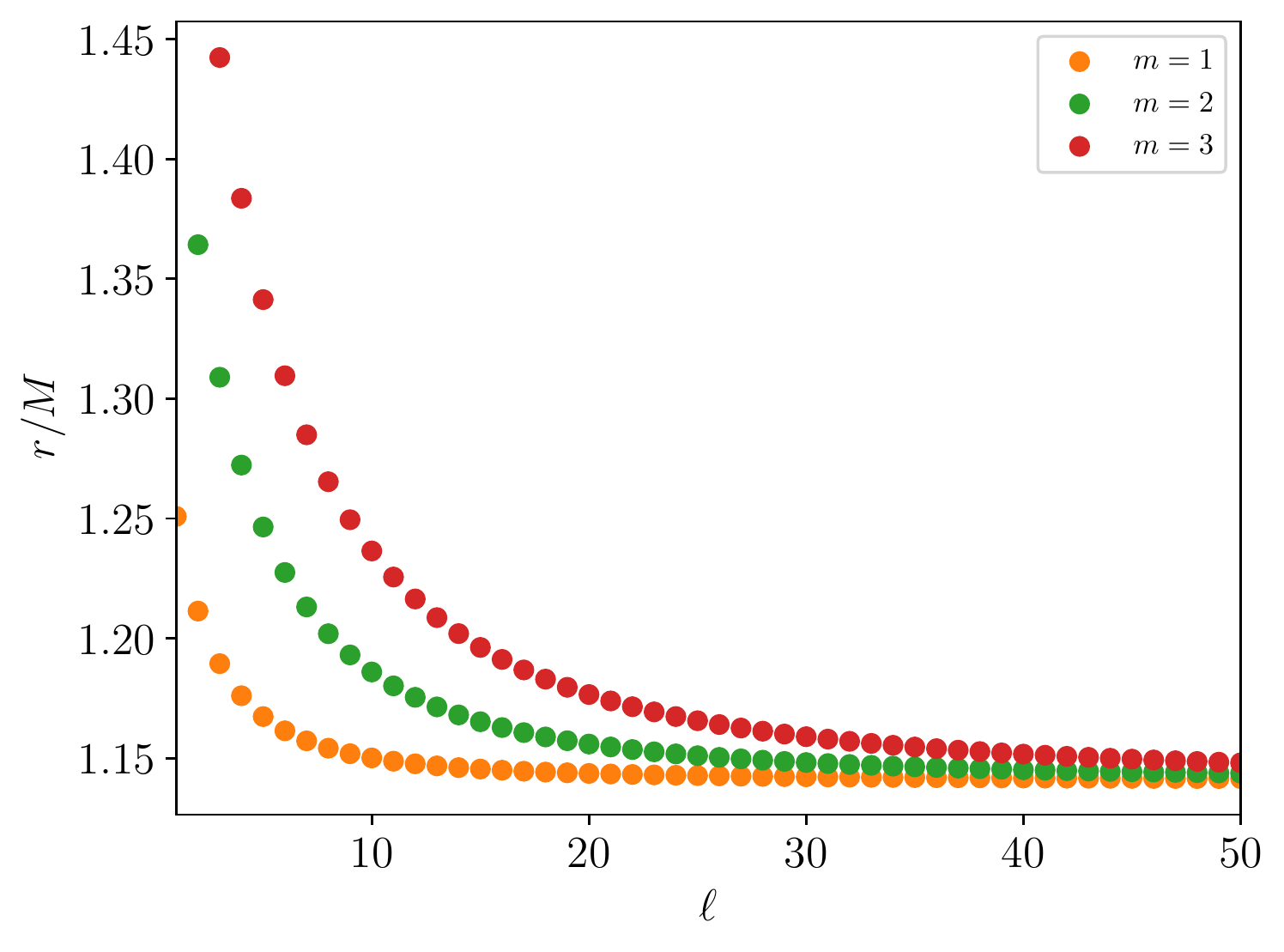}
  \caption{The surface locations of zero-frequency modes with $a = 0.99 M$ as a function of $\ell$ for different $m$. \label{fig:zeros_lm}}
\end{figure}

In this appendix, we discuss the behavior of $\ell > m$ modes. Fig.~\ref{fig:qnm_lm} shows some unstable modes with different $m$ and $\ell$. We can see that the $\ell = m$ modes are more unstable than the $\ell > m$ modes since their imaginary part is larger, thus their instability timescale is shorter. Furthermore, the $\ell > m$ modes become stable for smaller values of $\epsilon$, as highlighted by the dashed vertical lines.
Fig.~\ref{fig:zeros_lm} shows that the surface location of the zero-frequency mode always decreases as $\ell$ grows with fixed $m$. This means that modes with $\ell > m$ always become stable before modes with $\ell = m$ as the surface location $r_0$ increases. This ensures that when considering superradiance instability, we do not need to consider the modes with $\ell > m$.


\bibliography{references}

\end{document}